\documentclass[letterpaper,twocolumn]{article}

\usepackage[T1]{fontenc}
\usepackage{xeCJK}
\usepackage{xeCJK}
\usepackage{geometry}
\usepackage{setspace}
\usepackage[
doi=true,
articletitle=true,
maxauthors=3
]{achemso}

\usepackage{graphicx}
\usepackage{float}
\usepackage{multirow}
\usepackage{amssymb}
\usepackage{amsfonts}
\usepackage{hyperref}
\newfloat{scheme}{htbp}{los}
\floatname{scheme}{Scheme}
\floatname{chart}{Chart}
\newfloat{graph}{htbp}{loh}

\usepackage{chemformula} 
\usepackage[version = 4]{mhchem} 
\usepackage{amsmath}

\definecolor{myPurple}{RGB}{128,0,128}

\definecolor{myGreen}{RGB}{0,100,0}

\usepackage{comment}
\usepackage{authblk}
\author[1]{William Kai Alexander Worby}
\author[2]{Yuto Yokoyama}
\author[3]{Misa Kawaguchi}
\author[1]{Yoshiyuki Tagawa*}
\affil[1]{Dept. of Mechanical System Engineering, Tokyo University of Agriculture and Technology, Japan}
\affil[2]{Div. of Physical Science and Engineering, King Abdullah University of Science and Technology, Kingdom of Saudi Arabia}
\affil[3]{Institute of Engineering, Shinshu University, Japan}

\title{Apparent Breakdown of the Stress–Optic Rule in Rigid-Rod Suspensions:\\ A Stress-Partitioning Interpretation}

\date{*Email: tagawayo@cc.tuat.ac.jp}

\begin{document}

\maketitle

\begin{abstract}
A fundamental problem in the field of anisotropic macromolecular materials concerns how the flow-induced orientational order connects microscopic structure to macroscopic rheological stress and optical anisotropy. 
This study considered cellulose nanocrystal suspensions as rigid-rod model systems for the purpose of isolating orientational effects from chain stretching. 
Simultaneous rheo-optical measurements of birefringence, orientation angle, and shear stress were combined with a Fokker--Planck orientation model and stress decomposition. 
Our results show that, at low values of the Péclet number, birefringence and total stress exhibit proportionality similar to the stress–optic rule. However, the concentration dependence indicates that this proportionality reflects stress partitioning rather than a unique material coefficient. With increasing flow strength, orientational saturation is accompanied by a reduced relative Brownian stress contribution and a corresponding change in the total stress-based relation. 
Referencing the optical response to an estimated Brownian stress contribution yields a substantially more unified response across concentrations and flow conditions, supporting a stress-partitioning interpretation of apparent stress–optic rule breakdown.

\end{abstract}

\section{Keywords}
Stress--Optic Rule, Flow-Induced Orientation, Rheo-Optics, Stress-Partitioning, Colloidal Rods

\section{Introduction}
Understanding how flow-induced orientational order is connected to macroscopic rheological stress and optical anisotropy is a fundamental problem in anisotropic macromolecular and colloidal systems.
In many dilute or macroscopically isotropic complex fluids, Brownian motion randomizes the molecular or particle orientations at equilibrium.
As the fluid flows, however, the orientation distribution becomes anisotropic, giving rise to flow-induced birefringence\cite{Maxwell1874,Fuller1980,Mykhaylyk2016}.
Both the optical response and rheological stress are linked to the flow-induced microstructural state. Therefore, establishing their precise relationship is central to connecting microscopic orientation with macroscopic material response.

A classical framework for this connection is the stress-optic rule (SOR), which relates optical anisotropy to stress\cite{Philippoff1956,Wales1976,janeschitz-kriegl1983,Fuller1995}.
In polymeric systems, the SOR can arise when the same underlying molecular conformation governs the optical and mechanical responses.
Rheo-optical studies have examined flow-induced structural and orientational changes in polymeric systems\cite{Kume1997,Kadar2020,Sato2024,Muto2025b}.
For anisotropic particle suspensions, however, birefringence primarily reflects the anisotropy of the particle orientation distribution, whereas the experimentally measured stress may contain several contributions that are not directly associated with this orientational anisotropy.
This raises the fundamental question of what part of the macroscopic stress is actually represented by flow-induced birefringence in an anisotropic suspension.

The connection between stress and optical anisotropy provides a basis for mechanical measurements.
In solid materials, stress-induced optical anisotropy (or photoelasticity) has long been used for non-invasive visualization of internal stress fields\cite{Brewster1816,Aben2012,Ramesh2021}.
In complex fluids, flow-induced birefringence has developed into a rheo-optical method for probing structural dynamics and, when the SOR is applicable, for estimating stress or shear-rate fields.
Rheo-optical approaches have therefore been extended to flow visualization\cite{Sun1999,Hu2009}, quantitative flow and stress characterization\cite{Quinzani1995,Ober2011,Sun2016,Noto2020,Lane2023,Kim2017,Kobayashi2026}, and the application of stress-optical relations in three-dimensional and unsteady flows\cite{McAfee1974,Clemeur2004,Kusuno2025,Noto2025,Muto2025a,Kawaguchi2026}.

In applied rheo-optical studies, the SOR has been used to infer macroscopic stress from flow-induced birefringence\cite{Subramanian1996,Martyn2000}.
For the shear-flow geometry considered in this paper, we define the quantity as $q\equiv\Delta n\sin 2\chi/2$, where $\Delta n$ is the flow-induced birefringence and $\chi$ is the orientation angle.
The apparent SOR based on the total stress is then written as $q=C_{\rm app}\sigma_{\rm Total}$, where $C_{\rm app}$ is the apparent stress--optic coefficient defined with respect to the total stress.
The precise tensorial definitions of these quantities are given in the Theory section.
If $C_{\rm app}$ were independent of particle concentration and flow strength, the optical response would provide a direct measure of macroscopic stress.
Understanding why and under what conditions $C_{\rm app}$ varies is therefore essential for a physically grounded interpretation of rheo-optical measurements.

A simple macroscopic stress--birefringence relation, however, does not hold universally.
In polymeric systems, deviations from the SOR have been associated with finite chain extensibility, concentration and entanglement effects, associative interactions, and molecular-weight distribution\cite{Sridhar2000,Rothstein2002,Pellens2005,Luap2006}.
In micellar solutions, shear-induced structural changes can alter the proportionality between birefringence and total stress\cite{Shikata1994,Decruppe1995,Ito2016}, whereas 
in colloidal suspensions, birefringence and orientation angle vary with deformation rate and suspension conditions\cite{Peebles1964,Pindera1978,Detert2023}.
Rod-like colloidal systems exhibit nonlinear orientational responses and progressive particle alignment with increasing flow strength\cite{Lane2022,Calabrese2021}, raising the question of whether
a proportionality based on the total stress can remain unchanged beyond the linear-response regime.

Consequently, quantitative rheo-optical stress measurements often require system-specific calibration or stress-optic coefficients\cite{Sun2016,Lane2023,Nakamine2024,Kusuno2025,Habe2026}.
Moreover, stress-optic responses vary with concentration, flow conditions, and measurement geometry\cite{Salamon2020,Worby2024a,Kawaguchi2026}.
However, the physical origin of such variations, particularly which microscopic structures or stress components are represented by the optical response, remains unclear.
Resolving this issue is important for moving rheo-optics beyond empirical calibration toward a physically grounded interpretation of stress.

A key point is that birefringence does not directly reflect the total stress, but primarily the anisotropy of the particle orientation distribution.
When flow drives the orientation distribution away from equilibrium isotropy, birefringence develops, while the thermodynamic restoring tendency associated with Brownian rotational diffusion gives rise to the Brownian stress $\sigma_{\rm B}$.
Birefringence and Brownian stress can therefore be regarded as responses associated with the same orientational anisotropy.
Indeed, rheo-optical studies by Wagner and co-workers have shown that optical anisotropy in colloidal suspensions relates more directly to the thermodynamic contribution to stress than to the total stress\cite{Wagner1988,Bender1995,Bender1996}.
In the present rigid-rod framework, this orientational thermodynamic contribution is represented by $\sigma_{\rm B}$.
The corresponding Brownian-stress-based relation may therefore be written as $q=C_{\rm B}\sigma_{\rm B}$, where $C_{\rm B}$ is the stress-optic coefficient defined with respect to the Brownian stress.

Connecting this relation to the SOR based on the total stress is not straightforward.
The total stress $\sigma_{\rm Total}$ contains several physically distinct contributions, including solvent viscous, Brownian, excluded-volume, and hydrodynamic stresses\cite{Hinch1976,Fuller1995,Lang2019}.
Therefore, relating birefringence to the total stress requires a constitutive description that connects the orientational state to the individual stress components.

Orientation-statistical theories based on Fokker--Planck-type equations provide a framework for treating this correspondence quantitatively.
Such theories determine the orientation distribution under flow, from which the birefringence and orientational contributions to stress can be evaluated\cite{Hinch1976,Dhont2003}.
Although orientational information is essential for describing rod-like suspensions, the total stress cannot generally be determined from orientation alone, because additional interactions contribute to the rheological response \cite{Petrich2000,Salipante2025}.
In particular, the total-stress-based coefficient $C_{\rm app}$ differs from the Brownian-stress-based coefficient $C_{\rm B}$ when the relative contributions of Brownian and non-Brownian stresses change.

The central question addressed in this paper is therefore how the macroscopic SOR based on the total stress is related to the Brownian-stress-based optical relation.
Because both relations describe the same quantity $q$, they can be written as
$q=C_{\rm app}\sigma_{\rm Total}=C_{\rm B}\sigma_{\rm B}$, which gives
\begin{equation}
C_{\rm app}=C_{\rm B}
\frac{\sigma_{\rm B}}{\sigma_{\rm Total}}.
\label{eq:CovsCapp}
\end{equation}
This relation shows that $C_{\rm app}$ depends not only on the Brownian-stress-based coefficient $C_{\rm B}$, but also on the fraction of Brownian stress in the total stress.
Thus, changes in $C_{\rm app}$ do not necessarily imply a change in the underlying Brownian-stress-based relation, but may instead arise from changes in stress partitioning.

To organize this stress--birefringence correspondence, the relative
importance of flow-induced orientation and Brownian rotational relaxation
is characterized by the Péclet number, $Pe$.
For $Pe\ll1$, Brownian rotational diffusion dominates, and the flow-induced orientational anisotropy remains small; in this regime, birefringence exhibits an approximately linear response to the deformation rate\cite{Dhont2003}.
Around $Pe\sim O(1)$, flow-induced orientation becomes comparable to rotational diffusion, whereas at higher $Pe$, orientational nonlinearity and saturation become increasingly important.
$Pe$ therefore provides a natural parameter for distinguishing the linear-response regime from the nonlinear orientational regime.

In this study, cellulose nanocrystal suspensions were used as a model system of rigid rod-like particles.
Because the particle length remains essentially fixed, this system allows the coupling among orientational dynamics, optical anisotropy, and stress partitioning to be examined without additional nonlinearities associated with chain stretching or finite extensibility.
We simultaneously measured flow-induced birefringence, orientation angle, and shear stress using rheo-optical measurements, and compared the results with a Fokker--Planck-type orientation model and a constitutive stress model.
Through this approach, we were able to examine how the apparent SOR based on the total stress is related to the underlying orientation--stress correspondence, and how this relation changes as stress partitioning evolves with $Pe$.

\section{Theory}
\subsection{Stress--Optic Rule}
In general, the SOR describes a proportional relationship between the stress tensor and the refractive-index tensor \cite{Fuller1995}. 
As shown in Fig.~\ref{fig:coorinates}, we considered a two-dimensional Couette flow with the flow, velocity-gradient, and vorticity directions defined as the $x$-, $y$-, and $z$-axes, respectively. 
Let $\chi$ denote the angle between the principal axis of the refractive-index tensor in the $x$-$y$ plane and the flow direction. The SOR is then written as
\begin{equation}
\boldsymbol{\sigma}=\frac{1}{C}\mathbf{R}^{\rm T}(\chi)\boldsymbol{\Pi}\mathbf{R}(\chi),
\label{eq:SOR_tensor}
\end{equation}
where $C$ [Pa$^{-1}$] is the stress-optic coefficient, $\boldsymbol{\Pi}$ is the refractive-index tensor, and $\mathbf{R}(\chi)$ is the rotation matrix about the $z$-axis from the principal-axis system of $\boldsymbol{\Pi}$. 
The birefringence $\Delta n$, defined as the difference between the principal refractive indices, is related to the shear stress $\sigma_{xy}$ by
\begin{equation}
\sigma_{xy}=\frac{1}{2C}\Delta n \sin 2\chi.
\label{eq:SOR}
\end{equation}
In this study, $C$ was evaluated from Eq.~\eqref{eq:SOR} using the measured birefringence, orientation, and shear stress.

\subsection{Orientation Tensor}
\begin{figure}[tb!]
\centering
\includegraphics[width=0.7\linewidth]{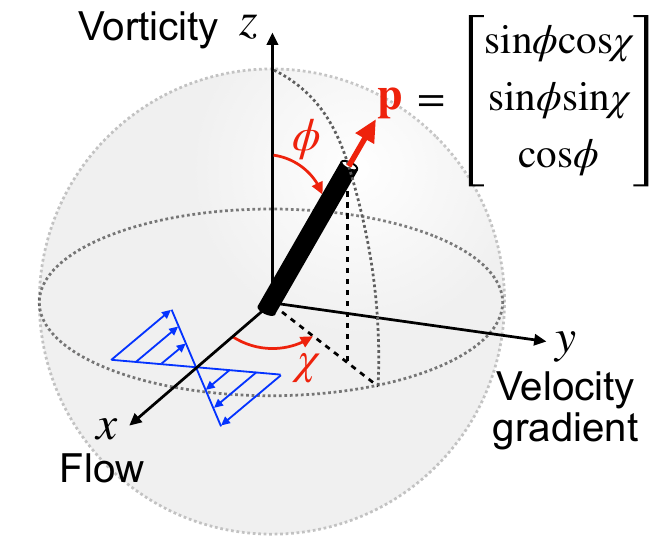}
\caption{\setstretch{0.8}Definition of the unit orientation vector $\boldsymbol{\rm p}$ for a rod-like particle and the angles $\phi$ and $\chi$. The shear flow is shown with the flow direction, velocity-gradient direction, and vorticity direction defined as the $x$-, $y$-, and $z$-axes, respectively.}
\label{fig:coorinates}
\end{figure}

As shown in Fig.~\ref{fig:coorinates}, the orientation of an individual rod-like particle is represented by the unit vector $\boldsymbol{\rm p}$ along its long axis:
\[
\boldsymbol{\rm p}
=
\begin{pmatrix}
p_x \\
p_y \\
p_z
\end{pmatrix}
=
\begin{pmatrix}
\sin\phi\cos\chi \\
\sin\phi\sin\chi \\
\cos\phi
\end{pmatrix}.
\]
Because rod-like particles are nonpolar, $\boldsymbol{\rm p}$ and $-\boldsymbol{\rm p}$ represent the same orientational state.

In a suspension, particle orientations are described by an orientation distribution function $f(\boldsymbol{\rm p},t)$ on the unit sphere, rather than by a single direction.
The quantity $f(\boldsymbol{\rm p},t)~\mathrm d\boldsymbol{\rm p}$ gives the probability that the orientation vector lies within the infinitesimal solid angle $\mathrm d\boldsymbol{\rm p}$ around $\boldsymbol{\rm p}$ at time $t$.
The distribution function can be normalized over the unit sphere as \cite{TuckerIII2022}
\[
\oint f(\boldsymbol{\rm p},t)~\mathrm d\boldsymbol{\rm p}
=
\int_{\chi = 0}^{2\pi}
\int_{\phi = 0}^{\pi}
f(\phi,\chi,t)\sin\phi~\mathrm d\phi\mathrm d\chi
=
1.
\]

The orientational state can be statistically characterized by the dyadic product $\boldsymbol{\rm p}\boldsymbol{\rm p}$:
\[
\boldsymbol{\rm p}\boldsymbol{\rm p}
=
\begin{pmatrix}
p_xp_x & p_xp_y & p_xp_z \\
p_yp_x & p_yp_y & p_yp_z \\
p_zp_x & p_zp_y & p_zp_z
\end{pmatrix}.
\]
The second-order orientation tensor $\boldsymbol{\rm S}$ is then defined as the orientational average of $\boldsymbol{\rm p}\boldsymbol{\rm p}$\cite{TuckerIII2022}:
\[
\boldsymbol{\rm S}
\equiv
\langle \boldsymbol{\rm p}\boldsymbol{\rm p} \rangle
=
\oint
\boldsymbol{\rm p}\boldsymbol{\rm p}
f(\boldsymbol{\rm p},t)~\mathrm d\boldsymbol{\rm p}.
\]
In an isotropic state, particles are oriented with equal probability in all directions, giving $\boldsymbol{\rm S}=\boldsymbol{\rm I}/3$, where $\boldsymbol{\rm I}$ is the identity tensor.
Under flow, the components of $\boldsymbol{\rm S}$ deviate from the isotropic state, and the tensor reflects the anisotropy induced by the flow field.

\subsection{Fokker--Planck-Type Equation}
To describe the orientational dynamics of rod-like particles in the dilute-to-semidilute regime, we used the Fokker--Planck-type orientation tensor model of Lang et al. \cite{Lang2019}.
In this model, the time evolution of the orientation tensor $\boldsymbol{\rm S}$ is governed by Brownian rotational diffusion, excluded-volume interactions, and flow-induced orientation.

Define the dimensionless time as $t^* = 6D_r^{\rm eff}t$, where $D_r^{\rm eff}$ [s$^{-1}$] is the effective rotational diffusion coefficient.
The effective Péclet number is then defined as
\begin{equation}
Pe = \frac{|\dot{\gamma}|}{D_r^{\rm eff}},
\label{eq:Peeff}
\end{equation}
where $\dot{\gamma}$ [s$^{-1}$] is the applied shear rate.
The nondimensional orientation tensor equation is given by \cite{Lang2019}
\begin{equation}
\begin{split}
\frac{\mathrm d\boldsymbol{\rm S}}{\mathrm d t^*}
={}
-\Bigl[
\boldsymbol{\rm S}
-\frac{1}{3}\boldsymbol{\rm I}
&+\frac{L}{d}\psi
\Bigl(
\boldsymbol{\rm S}^{(4)}:\boldsymbol{\rm S}
-\boldsymbol{\rm S}\cdot\boldsymbol{\rm S}
\Bigr)
\Bigr]
\\
&+
\frac{1}{6}Pe
\Bigl[
\hat{\boldsymbol{\rm D}}\cdot\boldsymbol{\rm S}
+\boldsymbol{\rm S}\cdot\hat{\boldsymbol{\rm D}}^{\rm T}
-2\boldsymbol{\rm S}^{(4)}:\hat{\boldsymbol{\rm E}}
\Bigr].
\end{split}
\label{eq:Lang_FP}
\end{equation}
Here, $\psi$ is the particle volume fraction and $L$ [m], $d$ [m] are the rod length and diameter, respectively.
The tensor $\boldsymbol{\rm D}$ is the velocity-gradient tensor, and $\boldsymbol{\rm E}=(\boldsymbol{\rm D}+\boldsymbol{\rm D}^{\rm T})/2$ is the rate-of-strain tensor.
The normalized tensors are defined as $\hat{\boldsymbol{\rm D}}=\boldsymbol{\rm D}/|\dot{\gamma}|$ and $\hat{\boldsymbol{\rm E}}=\boldsymbol{\rm E}/|\dot{\gamma}|$.
The first term on the right-hand side of Eq.~\eqref{eq:Lang_FP} represents relaxation toward the isotropic state and excluded-volume interactions, whereas the second term represents shear-induced orientation.
To close Eq.~\eqref{eq:Lang_FP}, we used the closure approximation for the fourth-order orientation tensor $\boldsymbol{\rm S}^{(4)}\equiv\langle {\bf SSSS}\rangle$, as proposed by Dhont and Briels \cite{Dhont2003}, following its implementation by Lang et al. \cite{Lang2019}.
For an arbitrary second-order tensor $\boldsymbol{\rm M}$,
\begin{equation}
\begin{split}
\boldsymbol{\rm S}^{(4)}:\boldsymbol{\rm M}
=
\dfrac{1}{5}\{\boldsymbol{\rm S}\cdot\overline{\boldsymbol{\rm M}}
+
\overline{\boldsymbol{\rm M}}&\cdot\boldsymbol{\rm S}
-
\boldsymbol{\rm S}\cdot\boldsymbol{\rm S}\cdot\overline{\boldsymbol{\rm M}}
-
\overline{\boldsymbol{\rm M}}\cdot\boldsymbol{\rm S}\cdot\boldsymbol{\rm S}
\\
&+
2\boldsymbol{\rm S}\cdot\overline{\boldsymbol{\rm M}}\cdot\boldsymbol{\rm S}
+
3\boldsymbol{\rm S}\boldsymbol{\rm S}:\overline{\boldsymbol{\rm M}}\}
\end{split}
\end{equation}
was used, where $\overline{\boldsymbol{\rm M}} =(\boldsymbol{\rm M}+ \boldsymbol{\rm M}^{\rm T})/2$.

In the dilute regime, the rotational diffusion coefficient of an isolated rod, $D_r^{0}$~[s$^{-1}$], is given by \cite{Doi1986}
\begin{equation}
D_r^{0}
=
\frac{3k_{\rm b}T{\rm ln}(L/d)}{\pi\eta_{\rm s}L^{3}}.
\label{eq:Dr0}
\end{equation}
Here, $k_{\rm b}$ [J/K] is the Boltzmann constant, $T$~[K] is the temperature, and $\eta_{\rm s}$ [Pa$\cdot$s] is the solvent viscosity.
In the semidilute regime, rotational diffusion is suppressed by interparticle interactions \cite{Doi1986,Tao2006}.
At high shear rates, however, flow-induced orientation reduces the geometrical constraints between rods.
To account for this effect, Lang et al.\cite{Lang2019} expressed the concentration- and orientation-dependent effective rotational diffusion coefficient as
\begin{equation}
D_r^{\rm eff}
=
\frac{
D_r^{0}
}{
1
+
\dfrac{1}{c}
\left[
\dfrac{5}{4}
\nu L^{3}
\bigg(
1
-
\dfrac{3}{5}
\boldsymbol{\rm S}:\boldsymbol{\rm S}
-
\dfrac{2}{5}
\left(
\boldsymbol{\rm S}:\boldsymbol{\rm S}
\right)^{2}
\bigg)
\right]^{2}
}.
\label{eq:Dr_eff_Lang}
\end{equation}
Here, $\nu$ [m$^{-3}$] is the particle number density and $c$ is a dimensionless parameter representing the strength of interparticle interactions.
In this study, we used $c\approx 1,300$, as reported by Teraoka et al.\cite{Teraoka1985}.
Equation~\eqref{eq:Dr_eff_Lang} gives $D_r^{\rm eff}\rightarrow D_r^{0}$ in the dilute and highly oriented (or high-$Pe$) limits, while describing the suppression of rotational diffusion under semidilute and weakly oriented conditions.

\subsection{Birefringence, Orientation Angle, and Stress}
The steady-state orientation tensor $\boldsymbol{\rm S}$ obtained by solving Eq.~\eqref{eq:Lang_FP} was used to evaluate the birefringence, orientation, and stress components.
Because birefringence is determined by the orientational anisotropy in the plane perpendicular to the optical path, the in-plane scalar order parameter $S$ was defined as \cite{Fuller1995}
\begin{equation}
S\equiv\sqrt{\left(S_{xx}-S_{yy}\right)^2+4S_{xy}^2}.
\label{eq:Birefringence}
\end{equation}
This quantity represents the magnitude of anisotropy of the in-plane orientation tensor.
$S=0$ for an isotropic state and $S=1$ for particles perfectly aligned in a single direction within the plane.

The calculated scalar order parameter was related to the experimentally measured birefringence using the intrinsic birefringence $\Delta n_0$ as\cite{Uetani2019}
\begin{equation}
\frac{\Delta n}{\psi}=S \Delta n_0 .
\label{eq:StoBire}
\end{equation}
Here, $\Delta n_0$ is the birefringence per unit volume fraction in the limit of perfect alignment and represents the material-dependent optical anisotropy.

The orientation angle $\chi$ was determined by the principal axes of the in-plane orientation tensor and was calculated as\cite{Fuller1995,Reddy2018}
\begin{equation}
\chi=
\frac{1}{2}
\tan^{-1}
\left(
\frac{2S_{xy}}{S_{xx}-S_{yy}}
\right).
\label{eq:Orientation}
\end{equation}
In comparisons with experimental results, $|\chi|$ was used to remove the sign dependence associated with the choice of coordinate system.

The suspension stress was evaluated from the orientation distribution using the constitutive equation proposed by Lang et al. \cite{Lang2019}.
The deviatoric stress tensor excluding the pressure term, $\boldsymbol{\sigma}_{}$, is expressed as the sum of the solvent viscous stress $\boldsymbol{\sigma}_{\rm S}$, Brownian stress $\boldsymbol{\sigma}_{\rm B}$, excluded-volume stress $\boldsymbol{\sigma}_{\rm EV}$, and hydrodynamic stress $\boldsymbol{\sigma}_{\rm H}$:
\begin{equation}
\begin{split}
\boldsymbol{\sigma}_{\rm Total}
&=
\boldsymbol{\sigma}_{\rm S}
+
\boldsymbol{\sigma}_{\rm B}
+
\boldsymbol{\sigma}_{\rm EV}
+
\boldsymbol{\sigma}_{\rm H},\\
\boldsymbol{\sigma}_{\rm S}
&=
2\eta_{\rm s}\dot{\gamma}\hat{\boldsymbol{\rm E}},\\
\boldsymbol{\sigma}_{\rm B}
&=
3\nu k_{\rm b}T
\left(
\boldsymbol{\rm S}
-
\dfrac{1}{3}\boldsymbol{\rm I}
\right),\\
\boldsymbol{\sigma}_{\rm EV}
&=
3\nu k_{\rm b}T
\frac{L\psi}{d}
\left(
\boldsymbol{\rm S}^{(4)}:\boldsymbol{\rm S}
-
\boldsymbol{\rm S}\cdot\boldsymbol{\rm S}
\right),\\
\boldsymbol{\sigma}_{\rm H}
&=
\frac{\nu k_{\rm b}T}{2}Pe^0
\left(
\boldsymbol{\rm S}^{(4)}:\hat{\boldsymbol{\rm E}}
-
\dfrac{1}{3}
\boldsymbol{\rm I}
\boldsymbol{\rm S}:\hat{\boldsymbol{\rm E}}
\right).
\end{split}
\label{eq:Lang_stress}
\end{equation}
Here, $Pe^{0}=|\dot{\gamma}|/D_r^{0}$ is the bare Péclet number based on the rotational diffusion coefficient of an isolated rod.
The stress compared with the rheometer measurements gives the shear component of the deviatoric stress tensor:
\begin{equation}
\sigma_{\rm Total}
\equiv
\left(\boldsymbol{\sigma}_{\rm Total}\right)_{xy}
=
\sigma_{\rm S}
+
\sigma_{\rm B}
+
\sigma_{\rm EV}
+
\sigma_{\rm H}.
\label{eq:stressdicompose}
\end{equation}
Each scalar stress $\sigma_i~(i={\rm S,B,EV,H})$ denotes the shear component of the corresponding stress tensor, $\sigma_i=(\boldsymbol{\sigma}_i)_{xy}$.
In this study, we substituted the steady-state orientation tensor $\boldsymbol{\rm S}$ obtained from Eq.~\eqref{eq:Lang_FP} into Eq.~\eqref{eq:Lang_stress} and evaluated the total and component stresses from their shear components.
Unless otherwise stated, $\sigma_{\rm Total}$ and $\sigma_i$ therefore refer to shear stress components; for example, $\sigma_{\rm B}/\sigma_{\rm Total}$ denotes $(\boldsymbol{\sigma}_{\rm B})_{xy}/(\boldsymbol{\sigma}_{\rm Total})_{xy}$.

\subsection{Stress--Optic Rule in Rigid-Rod Suspensions}
For rigid rods, the orientational state is described by the statistics of the unit orientation vector $\boldsymbol{\rm p}$.
As shown in Eq.~\eqref{eq:Birefringence}, flow-induced birefringence is determined by the anisotropy of the second-order orientation tensor $\boldsymbol{\rm S}$, while the Brownian stress in Eq.~\eqref{eq:Lang_stress} is directly related to the same tensor.
However, the total stress contains additional solvent viscous, excluded-volume, and hydrodynamic contributions, some of which involve the fourth-order orientation tensor $\boldsymbol{\rm S}^{(4)}$ \cite{Leal1972,Hinch1976,Fuller1995}.
Thus, a simple SOR based on the total stress is not generally applicable to rigid-rod suspensions.

Fuller \cite{Fuller1995} described a limiting case in which a simpler relation can emerge.
For strongly semidilute rigid-rod systems with $\nu L^3 \gg 1$ and under weak or moderate flow,
$Pe \lesssim O(1)$, the contribution involving the fourth-rank orientation tensor becomes asymptotically small.
If the solvent contribution is also negligible, an SOR-type proportionality can be recovered.
This result, however, represents a particular asymptotic limit, and does not imply that higher-order or non-Brownian stress contributions are generally negligible under low-$Pe$ conditions.
In the present analysis, by contrast, these non-Brownian stress contributions are explicitly retained.
At low values of $Pe$, the orientational response is approximately linear, and the relative stress contributions approach limiting values. Hence, an apparent SOR-type proportionality can emerge even when non-Brownian stresses remain finite.
As $Pe$ increases and the stress partition changes, the total-stress-based proportionality is no longer expected to remain constant.

\section{Material and Experiments}
\subsection{Suspension Preparation}\label{sec:SuspensionPrep}
Cellulose nanocrystals (CNC-HSFD, Cellulose Lab) were used as the dispersed particles.
CNC-HSFD (hereafter referred to as CNC) is a high-aspect-ratio rod-like crystalline material obtained by sulfuric acid hydrolysis of pulp.

A CNC stock suspension was prepared by dispersing a prescribed amount of CNCs in ultrapure water.
A stock suspension with a mass concentration of $6.0~\mathrm{wt.\%}$ was used in this study.
To improve CNC dispersibility, the stock suspension was sonicated using an ultrasonic homogenizer (UX-300, Mitsui Electric) in PWM mode at 40\% output, with 10 s ON/20 s OFF cycles for a total treatment time of 15 min and a net sonication time of 5 min.
Because heating during ultrasonication may affect the sulfate ester groups on the CNC surface \cite{Beck-Candanedo2005}, the sample container was cooled in ice water during the treatment.

The final CNC suspensions were prepared by diluting the freshly sonicated stock suspension in a Newtonian glycerol/water mixture solvent at a volume ratio of $90:10$ ($\eta_{\rm s}=241$~mPa$\cdot$s).
The CNC mass concentration was adjusted to $c_m=0.05$--$0.30~\mathrm{wt.\%}$, corresponding to a volume fraction of $\psi=0.04$--$0.25~\mathrm{vol.\%}$.
The shear viscosity of each CNC suspension was measured using a stress-controlled rheometer (MCR302, Anton Paar) equipped with a cone-plate geometry (CP50-0.5, Anton Paar).
Measurements were performed at $23~^\circ\mathrm{C}$.
The refractive index $\bar n$ and density $\rho$ [g/cm$^3$] of each suspension were also determined using an Abbe refractometer (ER-2S, Matsuyoshi Medical Instruments) and a pycnometer, respectively.

\subsection{Atomic Force Microscopy}\label{sec:AFM}
To evaluate the morphology and particle length distribution of the CNCs, atomic force microscopy (AFM) observations were performed.
AFM samples were prepared with reference to the Langmuir-Blodgett method\cite{Oliveira2022}.
The CNC stock suspension was sonicated and then diluted with ultrapure water to obtain a dilute CNC suspension with a concentration of approximately $1.0\times10^{-3}~\mathrm{wt.\%}$.
To promote CNC adsorption onto the mica surface, freshly cleaved mica substrates (50-D-12, NanoAndMore) were treated with an aqueous poly-L-lysine solution (0.1\% w/v, Sigma-Aldrich).
The lower half of each poly-L-lysine-treated mica substrate was then immersed in the dilute CNC suspension and withdrawn, leaving a thin film containing isolated CNCs on the substrate surface.

The prepared samples were observed using an atomic force microscope (AFM1000 NanoView, Suzhou FlyingMan Precision Instruments).
An aluminum-backside-coated cantilever (200AC-NA, OPUS by MikroMasch; nominal spring constant: $9~\mathrm{N/m}$; nominal tip radius: $7~\mathrm{nm}$) was used as the AFM probe.
AFM observations were performed in tapping mode, and the images were analyzed using Gwyddion.
Detailed sample preparation and measurement conditions are described in Appendix A.

\subsection{Rheo-Optical Measurements}
\subsubsection{Experimental Setup and Image Analysis}
\begin{figure}[tb!]
\centering
\includegraphics[width=1\linewidth]{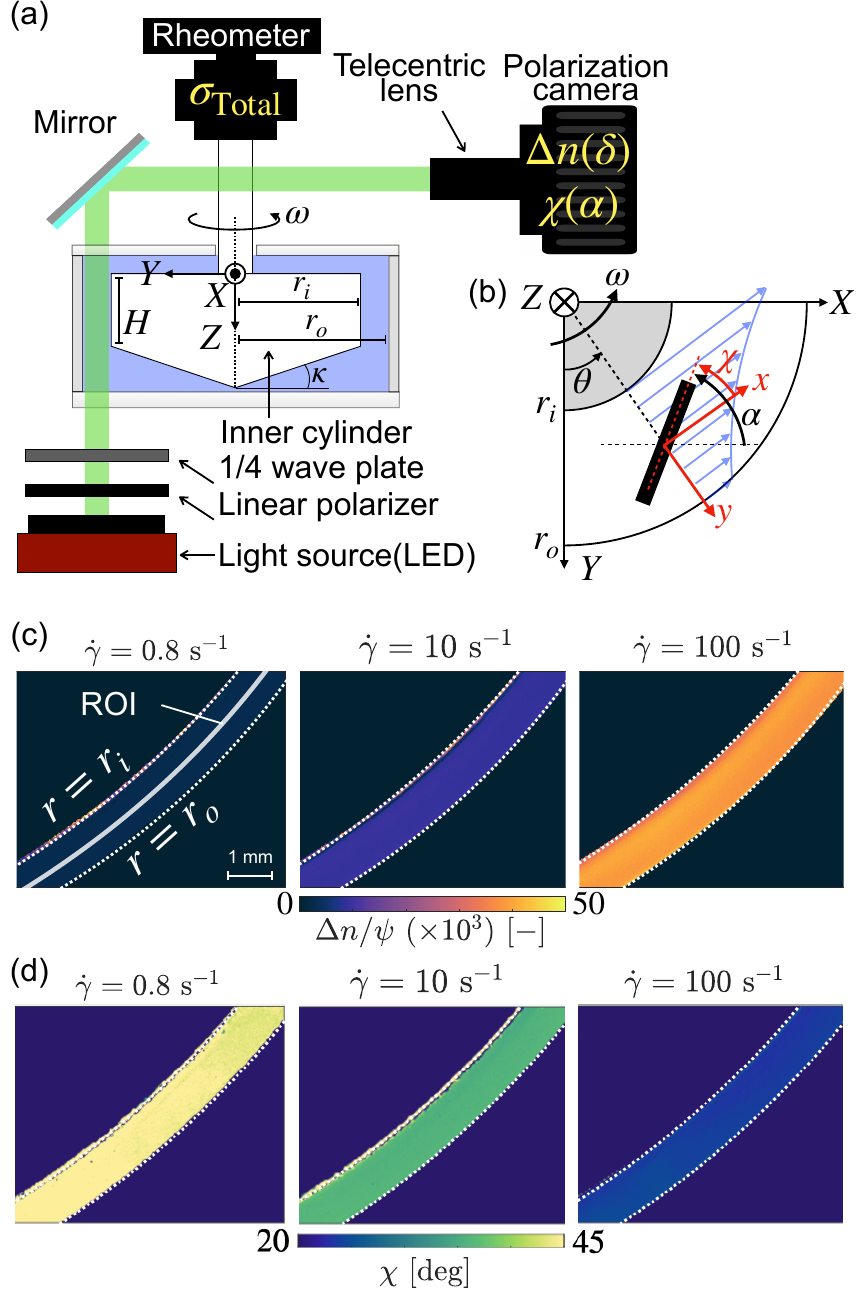}
\caption{\setstretch{0.8}(a) Schematic of the rheo-optical measurement system.
A Taylor--Couette flow cell was mounted on a stress-controlled rheometer, and the polarization state of the transmitted light was acquired using a polarization camera.
(b) Relationship between the laboratory coordinate system $(X,Y,Z)$ and the local coordinate system $(x,y,z)$ in the Taylor--Couette cell.
The orientation angle $\chi$ was calculated by subtracting the local flow-direction angle $\theta$ at each position.
Spatial distributions of (c) the concentration-normalized birefringence $\Delta n/\psi$ and (d) the orientation angle $\chi$ in the annular gap at a representative shear rate. The white dashed lines indicate the inner and outer cylinder walls; the ROI used for spatial averaging was set at the center of the gap.}
\label{fig:Setup}
\end{figure}

To perform rheo-optical measurements, a Taylor--Couette cell was mounted on a stress-controlled rheometer (MCR302, Anton Paar) together with a custom-made rotating inner cylinder (radius $r_i=18$ mm, height $H=10$ mm, taper angle $\kappa \approx18$ deg), as shown in Fig.~\ref{fig:Setup}(a).
The outer radius of the Taylor--Couette cell was $r_o=19$ mm.
Shear flow was applied to the CNC suspension by rotating the inner cylinder, and the total stress $\sigma_{\rm Total}$ was measured over a shear-rate range of $\dot{\gamma}=0.1$--$100~\mathrm{s^{-1}}$.
Simultaneously, the polarization state of light transmitted through the flow cell was obtained using a polarization camera (CRYSTA PI-5WP, Photron) equipped with a telecentric lens (VS-TCH1-65CO, VS Technology).
The birefringence $\Delta n$ and orientation angle $\chi$ with respect to the flow direction were evaluated from the polarization images.

A monochromatic LED light source with a wavelength of $\lambda=540~\mathrm{nm}$ (SOLIS-543C, Thorlabs) was used.
The emitted light was transmitted through a polarizer and a quarter-wave plate to generate circularly polarized light, which was incident on the CNC suspension in the Taylor--Couette cell.
The transmitted light became elliptically polarized owing to the optical anisotropy induced by the flow-induced orientation of the CNCs.
Polarization images were acquired at $424\times160$ pixels ($27~\mu\mathrm{m/pixel}$) and $500~\mathrm{fps}$.

The polarization camera integrates a wave-plate array with different optical-axis directions, a polarizer, and an image sensor, and calculates the polarization state from the intensities detected by four adjacent pixels \cite{Yoneyama2006,Yoneyama2007,Onuma2018}.
The obtained polarization state was analyzed on the Poincaré sphere.
The retardation $\delta$ [nm] and principal axis angle $\alpha$ [deg] in the laboratory coordinate system were obtained at each time and pixel using the control software (CRYSTA Stress Viewer, Photron).
For two-dimensional flow, the retardation and birefringence are related by $\delta=\Delta n H$.

The angle used to evaluate the stress--optic coefficient and compare the data with the orientation model is the orientation angle $\chi$ referenced to the local flow direction, rather than the principal axis angle $\alpha$ in the laboratory coordinate system.
Thus, denoting the local flow-direction angle at each position by $\theta$, $\chi$ was calculated as $\chi=\alpha-\theta$ (Fig.~\ref{fig:Setup}(b)).

Figures~\ref{fig:Setup}(c) and (d) show spatial distributions of the concentration-normalized birefringence $\Delta n/\psi$ and orientation angle $\chi$ with respect to the local flow direction under representative shear rate conditions, respectively.
Because no steep radial gradients were observed, representative values of $\Delta n$ and $\chi$ were evaluated within a region of interest (ROI) centered at the middle of the gap, $r=(r_i+r_o)/2$.
The ROI width was set to $3~\mathrm{pixels}$ ($\approx80~\mu\mathrm{m}$) in the radial direction.
At each shear rate, $\Delta n$ and $\chi$ were obtained by spatially averaging the data within the ROI.
Changing the ROI width induced only limited quantitative differences and did not affect the main conclusions.

The polarization camera measures retardation in the range $0\le \delta \le\lambda/2$.
Thus, phase wrapping occurs when the true retardation exceeds $\lambda/2$.
Assuming that the retardation varies continuously with time or shear rate, wrapping positions were identified from the measured retardation changes, and an unwrapping correction was applied.
This effect becomes more pronounced at higher concentrations or shear rates because $\Delta n$ increases with flow-induced orientation.
Phase wrapping also causes a discontinuous $90^\circ$ reversal in the principal axis angle $\alpha$. Thus, $\alpha$ was corrected based on the number of retardation wrapping events before conversion to $\chi$.
However, the correction of $\alpha$ is more susceptible to noise than the retardation correction, and scatter may remain in $\chi$, particularly in the high-$Pe$ regime.
Details of the measurement accuracy of the polarization camera and rheometer, as well as the phase wrapping correction, are provided in the Appendix B.

\subsubsection{Flow-Cessation Birefringence Relaxation}
Before evaluating $\Delta n$, $\chi$, and the stress-optic coefficient under steady flow, the effective rotational diffusion coefficient of the CNCs, $D_r^{\rm eff}$, was determined from the birefringence relaxation after flow cessation.
Once the flow ceases, oriented CNCs relax toward an isotropic state through Brownian rotational diffusion, causing the birefringence $\Delta n$ to decay with time.
For a monodisperse system, this relaxation is governed by a single rotational diffusion coefficient and follows a simple exponential decay \cite{Reddy2018}.
In contrast, real CNCs have a particle length distribution, and the rotational diffusion coefficient strongly depends on particle length.
Thus, the observed birefringence relaxation is expected to reflect a superposition of relaxation modes from particle populations with different lengths \cite{Tsvetkov2017,Brouzet2018,Arenas2018,Yokoyama2026}.

To describe this relaxation-time distribution phenomenologically, the birefringence decay was fitted with the Kohlrausch--Williams--Watts (KWW) function \cite{Alvarez1991}:
\begin{equation}
\frac{\Delta n(t)}{\Delta n(0)}=\exp\left[-\left(\frac{t}{\tau_{\rm kww}}\right)^{\beta}\right].
\label{eq:KWW}
\end{equation}
Here, $\tau_{\rm kww}$ is the characteristic time and $0<\beta\leq1$ is the stretching exponent, representing the breadth of the relaxation-time distribution.
The condition $\beta<1$ indicates a distribution of relaxation time scales, which is attributed to the length distribution of the CNCs.
Because the KWW function can be interpreted as a superposition of simple exponential relaxations, the mean relaxation time $\bar{\tau}$ is given by \cite{Alvarez1991}
\begin{equation}
\bar\tau=\frac{\tau_{\rm kww}}{\beta}\Gamma\left(\frac{1}{\beta}\right).
\label{eq:Tau}
\end{equation}
The effective rotational diffusion coefficient can then be evaluated as $D_r^{\rm eff}=1/(6\bar\tau)$ \cite{Fuller1995}.
The effective Péclet number was calculated from Eq.~\eqref{eq:Peeff} and used to organize the steady rheo-optical responses.

For the birefringence decay after flow cessation to be interpreted as CNC orientational relaxation, the flow decay time must be much shorter than the particle orientational relaxation time.
For planar Couette flow, the characteristic flow decay time is $\tau_f = h^2/\pi^2\mu$ \cite{batchelor1967}, where $h$ [m] is the gap width and $\mu$ [m$^2$/s] is the kinematic viscosity.
Under the present conditions, $\bar\tau/\tau_f \gtrsim O(10^2)$, indicating that the flow decays much faster than the CNC orientational relaxation.
Therefore, the observed birefringence decay was attributed to CNC orientational relaxation under conditions where the flow had effectively ceased.
A detailed evaluation is provided in the Appendix C.

\subsection{Simulations}
To evaluate the orientational response and stress components based on the Fokker--Planck equation, the orientation tensor of the rod-like particles was calculated numerically.
The axisymmetry of the Taylor--Couette flow was used, and the local flow field at the ROI center, $r=(r_i+r_o)/2$, was employed.
As shown in Fig.~\ref{fig:Setup}(b), the local coordinate system was defined such that the local $y$-axis at this position is collinear with the laboratory $Y$-axis ($\theta=0$), with the flow and velocity-gradient directions defined as the $x$- and $y$-axes, respectively.

When the outer cylinder is fixed and the inner cylinder rotates with an angular velocity of $\omega$ [rad/s], the azimuthal velocity $v_{\theta}$ [m/s] in the annular gap is given by \cite{Davey1962}
\[
v_{\theta}(r)
=
Ar
+
\frac{B}{r},
\]
where
\[
A
=
-\frac{r_i^2 \omega}{r_o^2-r_i^2},
\qquad
B
=
\frac{(r_ir_o)^2\omega}{r_o^2-r_i^2}.
\]
In the local coordinate system, where the $x$- and $y$-axes correspond to the azimuthal and radial directions, respectively, the velocity-gradient tensor $\boldsymbol{\rm D}$ is expressed as
\[
\boldsymbol{\rm D}
=
\begin{pmatrix}
2Bxy/r^4
&
-A-B(x^2-y^2)/r^4
&
0
\\
A-B(x^2-y^2)/r^4
&
-2Bxy/r^4
&
0
\\
0 & 0 & 0
\end{pmatrix}.
\]
The antisymmetric component containing $A$ corresponds to rigid-body rotation and does not contribute to the rate-of-strain tensor.
The components containing $B$ describe fluid deformation and determine the off-diagonal components of the rate-of-strain tensor $\boldsymbol{\rm E}$.
The local shear-rate magnitude at the ROI center was evaluated as $|\dot{\gamma}|=\sqrt{2\boldsymbol{\rm E}:\boldsymbol{\rm E}}$ and used to define $Pe$.

The time evolution of the orientation tensor in Eq.~\eqref{eq:Lang_FP} was solved numerically using MATLAB.
Because this equation contains nonlinear flow-orientation and higher-order tensor terms in addition to rotational relaxation, it can become stiff under certain calculation conditions.
Stable time integration was therefore performed using \texttt{ode15s}, an implicit backward-difference solver with variable order and step size.
The initial condition was the isotropic state, $\boldsymbol{\rm S}(0)=\boldsymbol{\rm I}/3$.
The solution was obtained over $t^*\in[0,15]$, and the steady-state values were used as theoretical predictions.
The resulting tensor $\boldsymbol{\rm S}$ was substituted into Eqs.~\eqref{eq:Birefringence}--\eqref{eq:Lang_stress} to calculate the birefringence, orientation angle, and stress components.
These theoretical values were then compared with the experimentally measured steady rheo-optical responses.

\section{Results and Discussion}
\subsection{Rheological and Morphological Characterization}\label{sec:AFMandViscosity}
\begin{figure}[tb!]
\centering
\includegraphics[width=1\linewidth]{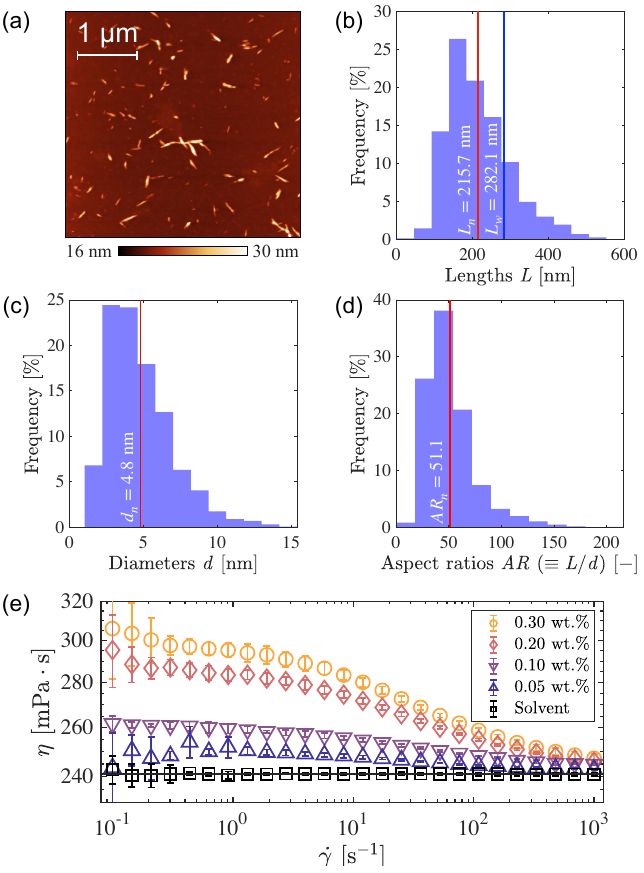}
\caption{\setstretch{0.8}(a) Representative AFM image. Size distributions of $n=1,604$ isolated CNC particles: (b) length $L$, (c) diameter $d$, and (d) aspect ratio $AR$. The red and blue lines indicate the number-average length $L_n$ and weight-average length $L_w$, respectively. (e) Flow curves of the CNC suspensions. Error bars represent the standard deviation from four measurements.}
\label{fig:AFM+Viscosity}
\end{figure}

Figure~\ref{fig:AFM+Viscosity}(a) shows a representative AFM image of the CNCs.
Particle dimensions were analyzed for isolated CNCs in the AFM images, and the distributions of length $L$, diameter $d$, and aspect ratio $AR$ were evaluated for $n=1,604$ particles.
The distributions of $L$, $d$, and $AR$ are shown in Fig.~\ref{fig:AFM+Viscosity}(b)--(d), respectively.
The CNCs exhibited a broad length distribution, confirming that the suspension used in this study was polydisperse.
The number-average length, diameter, and aspect ratio were $L_n = 216\pm88$ nm, $d_n = 4.8\pm2.2$ nm, and $AR_n = 51.1\pm23.7$, respectively.
These values are broadly consistent with previous reports\cite{Beck-Candanedo2005,Jakubek2018,Lane2023,Tanaka2015}.
Because rotational diffusion depends strongly on particle length ($D_r\propto L^{-3}$), the orientational dynamics are particularly sensitive to the longer-particle population\cite{Arenas2018,Yokoyama2026}.
In addition, birefringence measurements can preferentially weight
larger particle volumes in polydisperse suspensions\cite{Arenas2018,Tsvetkov2017}.
Therefore, in addition to $L_n$, the weight-average length $L_w$ was used as a practical representative length that emphasizes longer particles.
When $N_i$ particles have length $L_i$, we define 
\[
L_w
=
\frac{\sum_{i=1}^{n} L_i^2 N_i}
{\sum_{i=1}^{n} L_i N_i}.
\]
This definition gives greater weight to longer particles than the number-average length.
Applying this definition to the AFM length distribution yielded $L_w=282~\mathrm{nm}$.

The concentration regime of the CNC suspensions was examined based on Doi--Edwards theory \cite{Doi1986}.
For suspensions of rod-like particles with number density $\nu$ [m$^{-3}$], the dilute regime is defined by $\nu < 1/L^3$, whereas the semidilute regime is defined by $1/L^3\leq\nu\leq1/dL^{2}$.
Because this geometrical overlap criterion is defined for monodisperse rods, its application to polydisperse CNCs depends on the choice of representative length.

Using the dimensions of individual CNCs obtained from AFM, the effective number density $\nu_e$ was calculated following the method of Calabrese et al. \cite{Calabrese2021}.
The CNCs in the suspension were assumed to follow the particle volume distribution obtained from AFM.
Approximating each particle as a cylinder, $\nu_e$ was calculated from the average rod volume $V_{\rm rod}$ as
\[
\nu_e
=
\frac{\psi}{V_{\rm rod}}
=
\frac{\psi}
{\dfrac{1}{n}\sum_{i=1}^{n}\dfrac{\pi}{4}d_i^2L_i}.
\]

As listed in Table~\ref{tab:Quantities}, $\nu_e L_n^3\leq 1$ at lower concentrations when $L_n$ is used.
Thus, based on the Doi--Edwards overlap criterion, the CNC suspensions are located near the dilute-to-semidilute boundary.
In contrast, when $L_w$ is used, $\nu_e L_w^3>1$ for all concentrations.
This indicates that polydisperse CNC suspensions reach a weakly overlapped regime when the contribution of longer rods is included.
Therefore, the present suspensions should not be strictly classified as uniformly dilute or semidilute systems.
Rather, depending on the representative length, they span a regime from near the dilute-to-semidilute boundary to weakly overlapped conditions.

Figure~\ref{fig:AFM+Viscosity}(e) shows the steady shear viscosity of the prepared suspensions.
Even at the lowest concentration of $c_m = 0.05~\mathrm{wt.\%}$, shear thinning was observed.
This behavior reflects shear-induced orientation, which can occur even in dilute rod suspensions, and therefore does not by itself prove the contribution of interparticle interactions.
However, because the present concentration range satisfies the overlap criterion when evaluated using $L_w$, and because $D_r^{\rm eff}$ obtained from birefringence relaxation exhibits concentration dependence, the observed shear thinning may include weak interparticle interactions in addition to rod orientational response.
\begin{table}[tb!]
\centering
\caption{\setstretch{0.8}Characteristics of the CNC suspensions used in this study. Here, $c_m$ and $\psi$ denote the mass and volume fractions, respectively; $\rho$ is the suspension density; $\nu_e$ is the effective number density; $L_n$ and $L_w$ are the number- and weight-average lengths, respectively; and $\langle n \rangle$ is the suspension refractive index.}
\label{tab:Quantities}
\begin{tabular}{cccccc}\hline
\rule{0pt}{2.8ex}
$c_m~[\rm wt.\%]$ & $\psi~[\rm vol.\%]$ & $\rho~[\rm g/cm^3]$ & $\nu_e L_n^3$ & $\nu_e L_w^3$ & $\ \bar n$\\[0.4ex] \hline
0.30 & 0.247 & 1.238 & 4.53 & 10.1 &\multirow{4}{*}{1.458} \\
0.20 & 0.164 & 1.237 & 3.02 & 6.73\\
0.10 & 0.082 & 1.236 & 1.51 & 3.36\\
0.05 & 0.041 & 1.236 & 0.76 & 1.68\\ \hline
\end{tabular}
\end{table}

\subsection{Analysis of Alignment Dynamics}
\subsubsection{Flow Cessation}
\begin{figure}[tb!]
\centering
\includegraphics[width=1\linewidth]{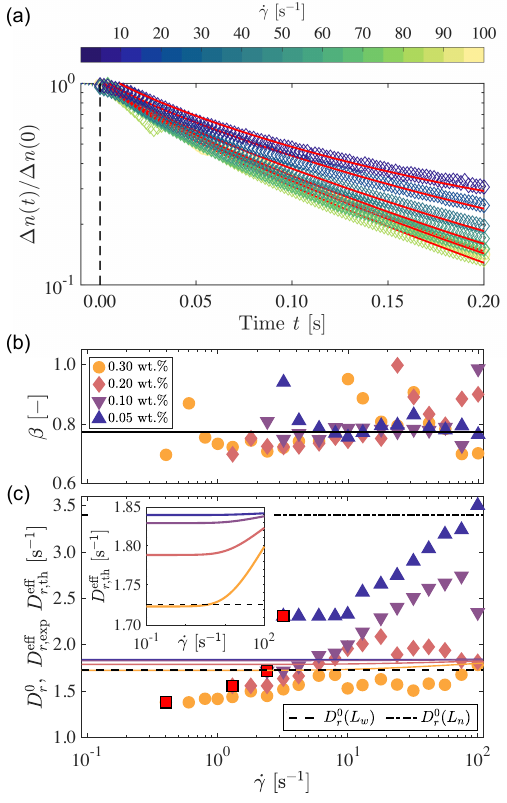}
\caption{\setstretch{0.8}Birefringence relaxation analysis after cessation of shear flow. (a) Temporal decay of the normalized birefringence $\Delta n(t)/\Delta n(0)$ at different shear rates $\dot{\gamma}$. The vertical dashed line indicates the time of flow cessation ($t=0$~s). Symbols represent experimental data for the CNC suspension with $c_m=0.20~\mathrm{wt.\%}$, and solid lines represent KWW fits obtained using Eq.~\eqref{eq:KWW}. The colors correspond to the shear rates applied before flow cessation. (b) Shear-rate dependence of the stretching exponent $\beta$ obtained from the KWW fits. The horizontal solid line indicates the average value of $\beta$ in the high-shear-rate regime. (c) Shear-rate dependence of the experimental effective rotational diffusion coefficient $D_{r,\mathrm{exp}}^{\mathrm{eff}}$ obtained from the mean relaxation time. The solid lines represent the theoretical effective rotational diffusion coefficient $D_{r,\mathrm{th}}^{\mathrm{eff}}$ calculated from Eq.~\eqref{eq:Dr_eff_Lang}; the inset shows an enlarged view. The red squares indicate the lowest shear-rate conditions for which reliable fitting was possible at each concentration. At lower shear rates, the corresponding red-square value at the same concentration was used as a representative value. The horizontal dashed and dash-dotted lines indicate $D_r^0(L_w)$ and $D_r^0(L_n)$, respectively.}\label{fig:Relaxation}
\end{figure}

Figure~\ref{fig:Relaxation}(a) shows the temporal decay of birefringence after flow cessation.
Each dataset was normalized by the value immediately after cessation, $\Delta n(0)$.
After cessation, flow-oriented CNCs relax toward an isotropic state through Brownian rotational diffusion, causing $\Delta n(t)$ to decrease with time.
This relaxation was used to evaluate the effective rotational diffusion coefficient $D_r^{\rm eff}$.

The birefringence relaxation was analyzed using the KWW function defined in Eqs.~\eqref{eq:KWW} and \eqref{eq:Tau}.
The data were fitted until $\Delta n(t)/\Delta n(0)$ reached 0.05.
Under low-shear-rate conditions, the initial birefringence was small, and $\Delta n(t)/\Delta n(0)$ did not always decay to 0.05.
For these conditions, $D_r^{\rm eff}$ was not evaluated individually by KWW fitting.
Instead, the value obtained at the lowest shear-rate condition for which reliable fitting was possible at the same concentration was used as a representative value.

Accordingly, Fig.~\ref{fig:Relaxation}(b) shows $\beta$ only for conditions with a sufficient relaxation range.
The value of $\beta$ approached a constant with increasing shear rate and became nearly independent of concentration in the high-shear-rate regime, giving $\beta = 0.77 \pm 0.01$.

Figure~\ref{fig:Relaxation}(c) shows the shear-rate dependence of the
effective rotational diffusion coefficient
$D_{r,\mathrm{exp}}^{\mathrm{eff}}$ obtained from birefringence relaxation.
The treatment of $D_{r,\mathrm{exp}}^{\mathrm{eff}}$ under low-shear-rate
conditions followed the procedure described above.
The value of $D_{r,\mathrm{exp}}^{\mathrm{eff}}$ decreased with increasing
concentration and increased with increasing shear rate.
The decrease at higher concentrations suggests that rod rotation is
constrained by interparticle interactions and geometrical hindrance.
In contrast, the increase with shear rate is consistent with the physical
picture for the dilute-to-semidilute transition regime, whereby
flow-induced orientation weakens the geometrical hindrance between rods
and the rotational diffusion approaches that of isolated rods
\cite{Doi1986,Tao2006}.

The obtained $D_{r,\mathrm{exp}}^{\mathrm{eff}}$ values were generally
located between the isolated-rod rotational diffusion coefficients
$D_r^0(L_w)$ and $D_r^0(L_n)$, evaluated from Eq.~\eqref{eq:Dr0}
using the weight-average length $L_w$ and number-average length $L_n$,
respectively.
In these evaluations, the effective diameter
$d_{\rm eff}=d_n+\delta d$ was used instead of the geometrical diameter
$d_n$ to account for the electric double layer on the CNC surface, where
$\delta d=22.6~\mathrm{nm}$ is the increase in diameter due to the electric
double layer \cite{Bertsch2019}.
As shown in the inset of Fig.~\ref{fig:Relaxation}(c), the theoretical
effective rotational diffusion coefficient
$D_{r,\mathrm{th}}^{\mathrm{eff}}$ obtained from
Eq.~\eqref{eq:Dr_eff_Lang} also increased with increasing shear rate.
However, this variation was smaller than that observed for
$D_{r,\mathrm{exp}}^{\mathrm{eff}}$, and the model did not reproduce the
stronger shear-rate dependence observed experimentally at lower
concentrations.

In polydisperse CNC suspensions, relaxation modes associated with different particle lengths do not contribute uniformly to the optical response because rotational diffusion is strongly length-dependent and larger particle volumes can contribute more strongly to birefringence
\cite{Tsvetkov2017,Brouzet2018,Arenas2018}.
Accordingly, $D_{r,\mathrm{exp}}^{\mathrm{eff}}$ should not be interpreted as the constant rotational diffusion coefficient of a single representative particle.
Rather, it should be regarded as an optically weighted effective relaxation rate that reflects both interparticle constraints and the particle-length populations contributing to the optical response.
In particular, shear-rate-dependent changes in optical weighting, as expected for polydisperse rod suspensions \cite{Yokoyama2026}, are not directly included in Eq.~\eqref{eq:Dr_eff_Lang}, which assumes
a single representative length.
This may contribute to the stronger shear-rate dependence of $D_{r,\mathrm{exp}}^{\mathrm{eff}}$ observed experimentally, particularly at lower concentrations.

\subsubsection{Steady State}
\begin{figure*}[tb!]
\centering
\includegraphics[width=1\linewidth]{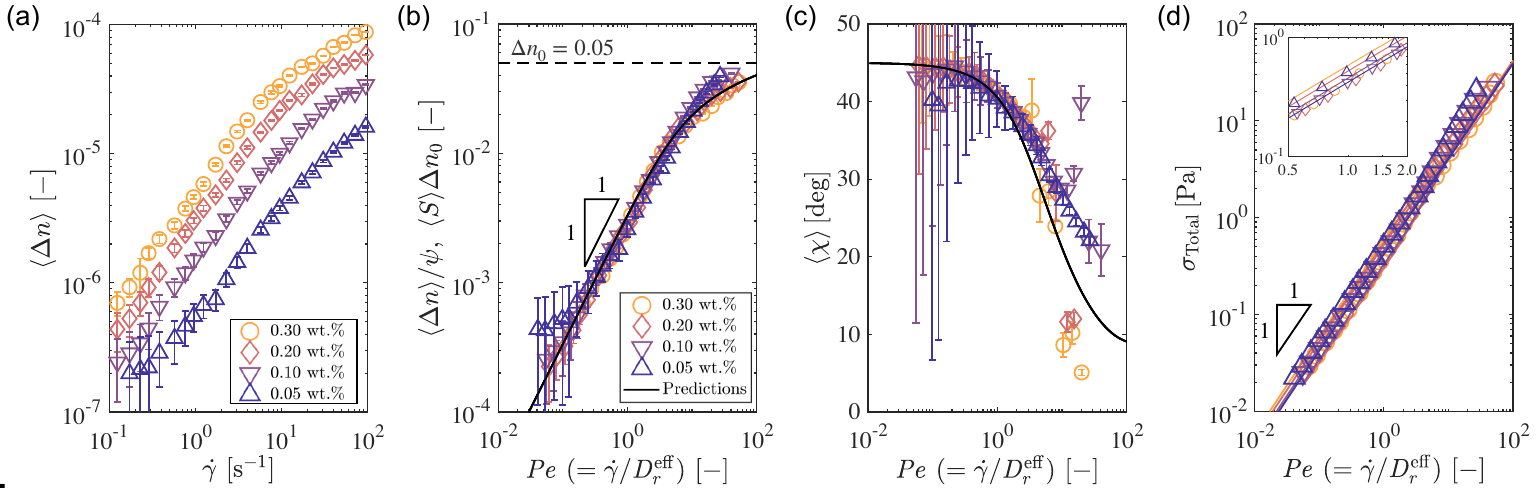}
    \caption{\setstretch{0.8}Steady rheo-optical responses as functions of $Pe$ based on the effective rotational diffusion coefficient.
    (a) Shear-rate dependence of the spatiotemporally averaged birefringence $\langle\Delta n\rangle$.
    $Pe$ dependence of (b) the volume fraction normalized birefringence $\langle\Delta n\rangle/\psi$, (c) the orientation angle $\langle\chi\rangle$, and (d) the total shear stress $\sigma_{\rm Total}$. Symbols represent experimental data.
    In panels (b)--(d), $Pe=|\dot{\gamma}|/D_{r,\rm exp}^{\rm eff}$ was evaluated using the concentration- and shear-rate-dependent $D_{r,\rm exp}^{\rm eff}$ shown in Fig.~\ref{fig:Relaxation}(c).
    Solid lines represent theoretical predictions obtained from the Fokker--Planck equation.
    In the theoretical calculations, $Pe=|\dot{\gamma}|/D_{r,\rm th}^{\rm eff}$ was defined using the theoretical effective rotational diffusion coefficient $D_{r,\rm th}^{\rm eff}$, which was determined self-consistently from the orientation tensor $\boldsymbol{\rm S}$ through Eq.~\eqref{eq:Dr_eff_Lang}.}\label{fig:Pe_vs_Δn_φ}
    \end{figure*}

Figure~\ref{fig:Pe_vs_Δn_φ}(a) shows the birefringence $\langle\Delta n\rangle$ measured at each suspension concentration and shear rate $\dot{\gamma}$.
Hereafter, $\langle \cdot \rangle$ denotes the spatial average over the ROI and the temporal average over the steady-state time window.
The error bars represent the spatial standard deviation within the ROI at each time, averaged over the same steady-state window.
In the low-shear-rate regime, $\langle\Delta n\rangle$ increased almost linearly with shear rate.
For $\dot{\gamma}\gtrsim10~\mathrm{s^{-1}}$, however, the increase became weaker, indicating a nonlinear orientational response of the particles.
This behavior arises because the CNCs orient toward the flow direction with increasing shear rate, and the degree of orientation approaches saturation.

Figure~\ref{fig:Pe_vs_Δn_φ}(b) shows the $Pe$ dependence of the volume-fraction-normalized birefringence, $\langle\Delta n\rangle/\psi$.
In the low-$Pe$ regime ($Pe\lesssim 1$), $\langle\Delta n\rangle/\psi$ increased linearly with $Pe$.
In the high-$Pe$ regime ($Pe\gtrsim O(1)$), the increase gradually weakened owing to the nonlinear development of particle orientation, and $\langle\Delta n\rangle/\psi$ tended to approach the intrinsic birefringence $\Delta n_0$.
At low values of $Pe$, the orientational relaxation through Brownian rotational diffusion is relatively strong compared with the flow-induced orientation, resulting in a small degree of orientation.
As $Pe$ increases, flow-induced orientation becomes dominant, and the CNCs increasingly align toward the flow direction.

Figure~\ref{fig:Pe_vs_Δn_φ}(c) shows the $Pe$ dependence of the orientation angle $\langle\chi\rangle$ with respect to the flow direction.
The orientation angle decreased monotonically from $45^\circ$ toward $0^\circ$ with increasing $Pe$, indicating that the particle long axes progressively oriented toward the flow direction.
The scatter in $\langle\chi\rangle$ observed in the high-$Pe$ regime for $c_m = 0.20$ and $0.30~\mathrm{wt.\%}$ can be partly attributed to uncertainty associated with the phase wrapping correction described above.

Figure~\ref{fig:Pe_vs_Δn_φ}(d) shows the $Pe$ dependence of the total stress $\sigma_{\rm Total}$ measured by the rheometer.
Whereas $\langle\Delta n\rangle/\psi$ showed a linear response at low values of $Pe$ and a saturating nonlinear response under high-$Pe$ conditions, $\sigma_{\rm Total}$ continued to increase over the entire $Pe$ range.
Thus, in the high-$Pe$ regime, the birefringence response weakened owing to orientational saturation, whereas the total stress continued to increase because it includes both solvent viscous stress and particle-derived stress components.
In this study, $\sigma_{\rm Total}$ was used as the reference stress for evaluating the apparent stress--optic coefficient $C_{\rm app}$ and the normalized stress discussed below.

The solid lines in Fig.~\ref{fig:Pe_vs_Δn_φ}(b)--\ref{fig:Pe_vs_Δn_φ}(d) represent theoretical predictions based on the Fokker--Planck-type orientation model.
For the experimental data, $Pe$ was calculated using $D_{r,\mathrm{exp}}^{\mathrm{eff}}$, as described in the previous section.
In the theoretical calculations, $D_r^0$ was evaluated using the weight-average length $L_w$ and effective diameter $d_{\rm eff}$, and $D_{r,\mathrm{th}}^{\mathrm{eff}}$ was determined self-consistently with the orientation tensor $\boldsymbol{\rm S}$ using Eq.~\eqref{eq:Dr_eff_Lang}.

For both the experiments and theory, the concentration- and shear-rate-dependent orientational relaxation is represented by the effective rotational diffusion coefficient $D_r^{\rm eff}$.
Consequently, $\langle\Delta n\rangle/\psi$ and $\langle\chi\rangle$ collapsed onto master curves as functions of $Pe=|\dot{\gamma}|/D_r^{\rm eff}$ over the concentration range examined in this study.

As shown in Fig.~\ref{fig:Pe_vs_Δn_φ}(b), the theoretical calculation broadly reproduced the increase in $\langle\Delta n\rangle/\psi$ and its tendency toward saturation in the high-$Pe$ regime.
This indicates that the magnitude of the in-plane orientational anisotropy can be described up to high values of $Pe$ using an equivalent-rod model with a single representative length.
The intrinsic birefringence used in the theoretical calculation was $\Delta n_0=0.050\pm0.001$, which lies within the range of previously reported values \cite{Uetani2019,Khan2024}.

As shown in Fig.~\ref{fig:Pe_vs_Δn_φ}(c), the theoretical calculation also broadly reproduced the decrease in $\langle\chi\rangle$ with increasing $Pe$.
In the high-$Pe$ regime, however, the experiments tended to produce larger $\langle\chi\rangle$ values than the theoretical prediction, even considering the scatter associated with the phase wrapping correction.
This suggests that orientation toward the flow direction progresses more gradually in the experiments than predicted by the theory.
A similar shift in the orientational response toward higher $Pe$ values relative to theoretical predictions has been reported in other suspension systems \cite{Calabrese2022,Reddy2018}.

The model used in the present study assumes a rod system with a single representative length and does not explicitly describe the orientational response of each particle length population in the polydisperse CNC suspension.
For the experimental data, $Pe$ was calculated using $D_{r,\mathrm{exp}}^{\mathrm{eff}}$, as obtained in the previous section.
Thus, the broad agreement between theory and experiment does not imply that the polydisperse system can be rigorously described as a monodisperse system.
Rather, it indicates that the averaged orientational response can be approximately organized as that of an equivalent rod system characterized by a single effective relaxation time.

The present model represents the polydisperse CNC suspension by a single effective rod length.
This approximation cannot capture length-dependent orientation\cite{Brouzet2018,Yokoyama2026,mrecktenwald2026} or the potentially different weighting of the particle-length distribution in the optical and rheological responses \cite{Arenas2018,Tanaka2017}.
Thus, the particle populations contributing most strongly to birefringence may not coincide with those contributing most strongly to the rheological stress.
Such differences may contribute to the remaining deviations between experiment and the single-length model, particularly at high $Pe$.
A fully polydisperse description would therefore require separate averaging of the length-resolved optical and stress responses over the particle distribution.

The total stress $\sigma_{\rm Total}$ includes several CNC-derived stress components in addition to the solvent viscous stress.
Moreover, the conversion from $Pe$ to shear rate, $\dot{\gamma}=PeD_r^{\mathrm{eff}}$, varies with the concentration-dependent effective rotational diffusion coefficient.
Therefore, unlike $\langle\Delta n\rangle/\psi$ and $\langle\chi\rangle$, $\sigma_{\rm Total}$ retains concentration dependence and does not collapse onto a single master curve.
As shown in the inset of Fig.~\ref{fig:Pe_vs_Δn_φ}(d), the theoretical curves exhibit concentration-dependent responses.
Over the present measurement range, the theoretical calculation tended to overestimate the absolute value of $\sigma_{\rm Total}$ compared with the experimental data.
This discrepancy affects the evaluation of the Brownian stress contribution to the total stress, as discussed below.
As described in Appendix B, validation experiments using silicone oils confirmed that the stress measured with the custom rheometer includes systematic uncertainty in its absolute value.
This uncertainty mainly affects comparisons of absolute stress values.
Its influence on relative trends in the $Pe$ and concentration dependences is considered smaller because these trends were obtained using the same instrument and analysis procedure.

Taken together, the main rheo-optical trends observed over the present measurement range could broadly be explained by the orientation model for rigid rods.
Specifically, the saturation of $\langle\Delta n\rangle/\psi$, the decrease in $\langle\chi\rangle$, and the increase in $\sigma_{\rm Total}$ can be described as the flow-induced orientational responses of rod-like particles, without assuming the formation of nematic-like order.

\subsection{Evaluation of Stress--Optic Coefficient}
\subsubsection{Apparent Stress--Optic Coefficient}
\begin{figure*}[tb!]
\centering
\includegraphics[width=1\linewidth]{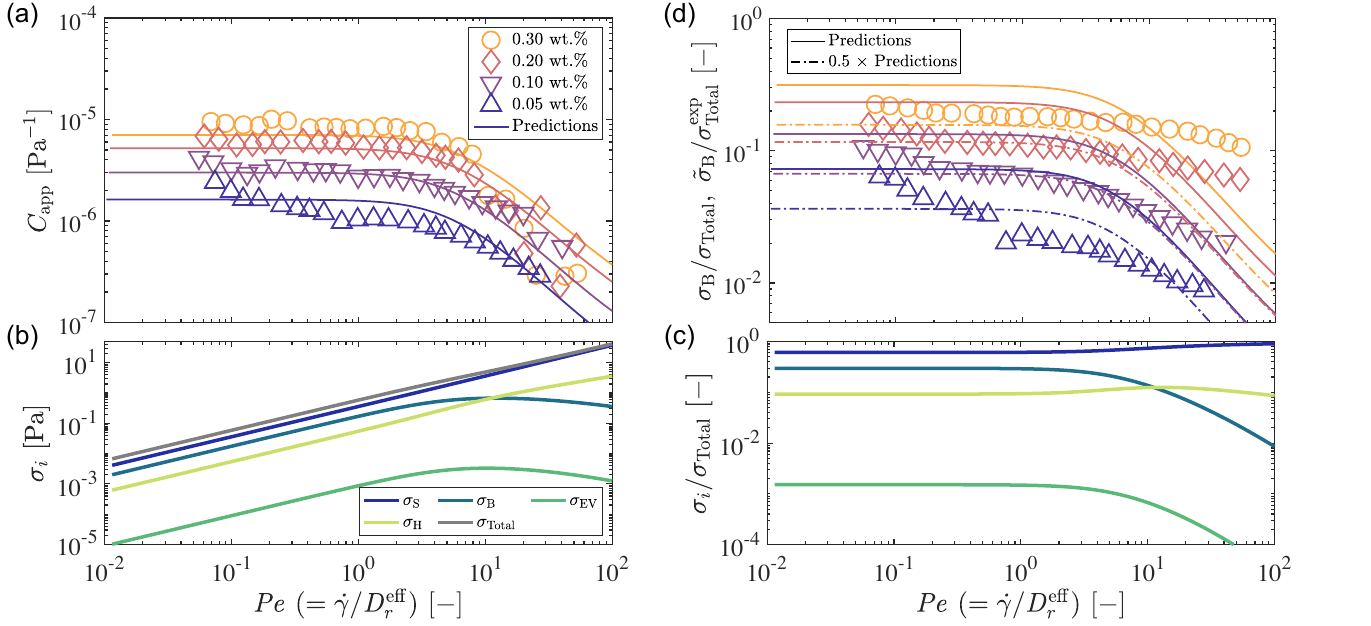}
\caption{\setstretch{0.8}(a) $Pe$ dependence of the apparent stress--optic coefficient $C_{\rm app}$.
    Symbols represent experimental data at each concentration, and the solid line represents the theoretical prediction based on the Fokker--Planck equation.
    (b) Absolute values of the stress components $\sigma_{\rm S}$, $\sigma_{\rm B}$, $\sigma_{\rm EV}$, and $\sigma_{\rm H}$ calculated using the constitutive equation of Lang et al. \cite{Lang2019}.
    (c) Relative contributions of these stress components normalized by the total stress $\sigma_{\rm Total}$.
    Panels (b) and (c) show the results for $c_m=0.30~\mathrm{wt.\%}$ as a representative example.
    In the high-$Pe$ regime, the relative Brownian stress contribution decreases, corresponding to the apparent decrease in $C_{\rm app}$ defined with respect to the total stress.
    (d) $Pe$ dependence of the Brownian stress contribution to the total stress.
    For the experimental data, the estimated value $\tilde{\sigma}_{\rm B}/\sigma_{\rm Total}^{\rm exp}$ is shown.
    Solid lines represent the theoretical $\sigma_{\rm B}/\sigma_{\rm Total}$, and dotted lines indicate $0.5$ times the theoretical prediction as a guide to the eye.}
    \label{fig:Capp}
    \end{figure*}

Figure~\ref{fig:Capp}(a) shows the apparent stress--optic coefficient $C_{\rm app}(Pe)$ evaluated at each value of $Pe$.
Here, $C_{\rm app}$ was calculated from Eq.~\eqref{eq:SOR} as the ratio between the polarization-derived quantity and the total stress under each flow condition:
\begin{equation}
C_{\rm app}(Pe)=\frac{\Delta n \sin 2\chi}{2\sigma_{\rm Total}} .
\label{eq:SOC}
\end{equation}
For the experimental values, the rheometer-measured total stress, $\sigma_{\rm Total}^{\rm exp}$, was used.
For the theoretical values, $C_{\rm app}$ was calculated using $\Delta n$, $\chi$, and $\sigma_{\rm Total}$ obtained from the Fokker--Planck model.

In the low-$Pe$ regime ($Pe\lesssim 1$), $C_{\rm app}$ remained nearly constant at each concentration.
This behavior indicates an apparent SOR-type relationship between birefringence and total stress in this regime.
However, the plateau value of $C_{\rm app}$ depended on concentration, showing that $C_{\rm app}$ defined with respect to the total stress is not a single material constant.
Because the Brownian stress increases with particle number density $\nu$, its fraction in the total stress also varies with concentration.
Based on Eq.~\eqref{eq:CovsCapp}, the concentration dependence of $C_{\rm app}$ in the low-$Pe$ regime can therefore be interpreted as a difference in the Brownian stress fraction
$\sigma_{\rm B}/\sigma_{\rm Total}$, rather than a change in the underlying Brownian-stress-based relation.
To examine the origin of the decrease in $C_{\rm app}$ with increasing $Pe$, the stress components constituting $\sigma_{\rm Total}$ were evaluated using Eq.~\eqref{eq:stressdicompose}.
As a representative example, the absolute values of the individual stress components and their relative contributions to the total stress at $c_m = 0.30~\mathrm{wt.\%}$ are shown in Fig.~\ref{fig:Capp}(b) and \ref{fig:Capp}(c), respectively.
In the low-$Pe$ regime, all stress components increased linearly with $Pe$, and the Brownian stress fraction remained nearly constant.
Correspondingly, $C_{\rm app}$ exhibited a plateau at each concentration.
In the high-$Pe$ regime, the rods became strongly oriented along the flow direction, leading to orientational saturation, while $\chi$ asymptotically approached $0^\circ$.
As a result, the increase in $S_{xy}=\langle p_xp_y\rangle$, which governs the shear component of the Brownian stress, became weaker.
Consequently, the contribution of $\sigma_{\rm B}$ to the total stress decreased, whereas the relative contributions of the solvent viscous stress and hydrodynamic stress increased.
Thus, the $Pe$ dependence of $C_{\rm app}$ can be understood as a change in the Brownian stress fraction relative to the total stress, rather than a change in the absolute Brownian stress itself.

In the following, the experimental results are organized from the viewpoint of this stress-component ratio.
Figure~\ref{fig:Capp}(d) shows the $Pe$ dependence of the Brownian stress contribution to the total stress.
The stress components directly accessible from experiments are the rheometer-measured total stress, $\sigma_{\rm Total}^{\rm exp}$, and the solvent viscous stress, $\sigma_{\rm S}=\eta_{\rm s}\dot{\gamma}$, calculated from the independently measured solvent viscosity.
Therefore, $\sigma_{\rm Total}^{\rm exp}-\eta_{\rm s}\dot{\gamma}$ represents the experimentally obtained particle-derived stress.
Because this quantity includes the Brownian stress $\sigma_{\rm B}$, excluded-volume stress $\sigma_{\rm EV}$, and hydrodynamic stress $\sigma_{\rm H}$, it cannot be directly identified with $\sigma_{\rm B}$.

In the theoretical calculation based on the constitutive equation of Lang et al. \cite{Lang2019}, 
$\sigma_{\rm EV}$ was 2--3 orders of magnitude smaller than the other stress components 
(Fig.~\ref{fig:Capp}(c)).
In the low-$Pe$ limit, linearization of the constitutive equation of Lang et al. \cite{Lang2019} gives the stress partitioning
\begin{equation}
\frac{\sigma_{\rm H}}
{\sigma_{\rm B}+\sigma_{\rm EV}}
\simeq
\frac{1}{3}
\frac{D_r^{\rm eff}}{D_r^{0}}.
\label{eq:sigmaH_sigmaBEV}
\end{equation}
Because $\sigma_{\rm EV}$ is negligibly small compared with $\sigma_{\rm B}$ in the present system, this relation can be approximated as
\begin{equation}
\frac{\sigma_{\rm H}}{\sigma_{\rm B}}
\simeq
\frac{1}{3}
\frac{D_r^{\rm eff}}{D_r^{0}}.
\label{eq:sigmaBsigmaH}
\end{equation}
Accordingly, using
$\sigma_{\rm Total}^{\rm exp}-\eta_{\rm s}\dot{\gamma}
\simeq
\sigma_{\rm B}+\sigma_{\rm H}$,
the Brownian stress can be estimated as
\begin{equation}
\tilde{\sigma}_{\rm B}
\approx
\frac{\sigma_{\rm Total}^{\rm exp}-\eta_{\rm s}\dot{\gamma}}
{1+\dfrac{1}{3}\dfrac{D_r^{\rm eff}}{D_r^{0}}}.
\label{eq:sigmaB_estimated}
\end{equation}
Here, $\tilde{\sigma}_{\rm B}$ is not the Brownian stress directly separated from the experiment, but a model-dependent estimate obtained by decomposing the measured particle-derived stress.
This decomposition relies on both the negligible contribution of $\sigma_{\rm EV}$ and the low-$Pe$ linearized stress partitioning derived from the constitutive equation.
In the following, this estimated value is denoted by $\tilde{\sigma}_{\rm B}$ to distinguish it from the theoretical Brownian stress $\sigma_{\rm B}$.

The absolute value of $\tilde{\sigma}_{\rm B}$ is subject to uncertainties associated with the representative particle length through $D_r^{0}$, as well as with the low-$Pe$ and weak-orientation approximations underlying the stress partitioning.
Therefore, for $Pe\gtrsim O(1)$, $\tilde{\sigma}_{\rm B}$ does not necessarily provide a quantitative measure of the true Brownian stress.
In the high-$Pe$ regime, $\tilde{\sigma}_{\rm B}$ is thus treated as an indicator obtained by extrapolating the low-$Pe$ stress partitioning, and the discussion focuses on its trend rather than its absolute magnitude.

Figure~\ref{fig:Capp}(d) shows the $Pe$ dependence of the estimated Brownian stress fraction, $\tilde{\sigma}_{\rm B}/\sigma_{\rm Total}^{\rm exp}$, obtained from Eq.~\eqref{eq:sigmaB_estimated}.
In the low-$Pe$ regime, $\tilde{\sigma}_{\rm B}/\sigma_{\rm Total}^{\rm exp}$ remained nearly constant, corresponding to the plateau in $C_{\rm app}$.
However, its absolute value was approximately half of $\sigma_{\rm B}/\sigma_{\rm Total}$ obtained from the Fokker--Planck model.
This systematic difference can be partly explained by the difference in the absolute total stress between the experiments and theory.
In the present system, the solvent viscous stress $\eta_{\rm s}\dot{\gamma}$ accounted for a large fraction of the total stress.
Therefore, when $\sigma_{\rm Total}^{\rm exp}$ was less than the theoretical value (Fig.~\ref{fig:Pe_vs_Δn_φ}(d)), the particle-derived stress obtained after subtracting the solvent contribution, $\sigma_{\rm Total}^{\rm exp}-\eta_{\rm s}\dot{\gamma}$, decreased more substantially in relative terms.
As a result, both $\tilde{\sigma}_{\rm B}$ and $\tilde{\sigma}_{\rm B}/\sigma_{\rm Total}^{\rm exp}$ were evaluated to be smaller than the theoretical values.

Thus, the difference in the absolute values between experiment and theory cannot be interpreted solely as a difference in the actual Brownian stress contribution.
Nevertheless, the near-constant estimated Brownian stress fraction in the low-$Pe$ regime is consistent with the theoretical picture that the plateau in $C_{\rm app}$ corresponds to a regime in which the Brownian stress contribution is approximately constant.

\subsubsection{Brownian-Stress-Based Stress--Optic Relation}
\begin{figure}[tb!]
\centering
\includegraphics[width=1\linewidth]{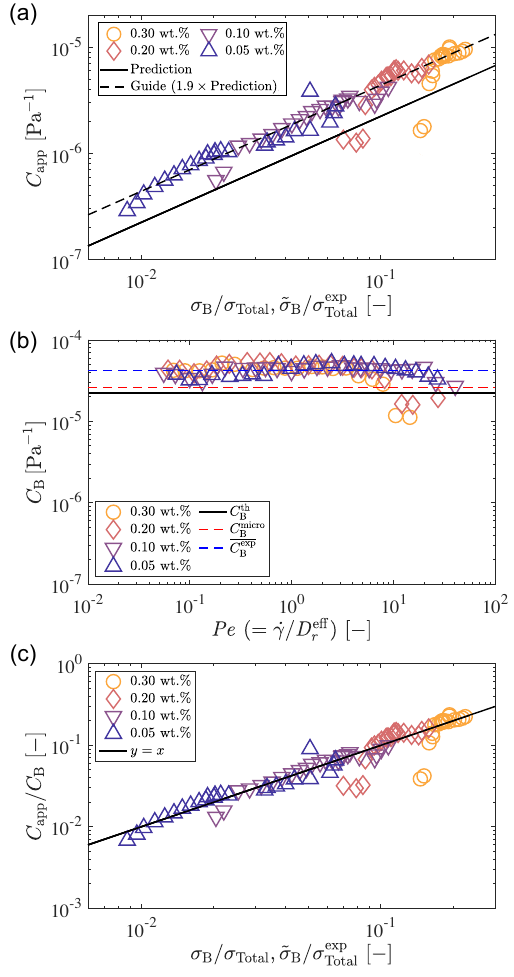}
    \caption{\setstretch{0.8}Stress--birefringence correspondence based on the Brownian stress contribution.
    (a) Relationship between the apparent stress--optic coefficient $C_{\mathrm{app}}$ and the Brownian stress fraction.
    Experimental values used $\tilde{\sigma}_{\mathrm{B}}/\sigma_{\mathrm{Total}}^{\mathrm{exp}}$ estimated from Eq.~\eqref{eq:sigmaB_estimated}, whereas theoretical values used $\sigma_{\mathrm{B}}/\sigma_{\mathrm{Total}}$ obtained from the Fokker--Planck model and the constitutive equation of Lang et al. \cite{Lang2019}.
    The dashed line is a guide obtained by multiplying the theoretical prediction by 1.9 and is not a fit.
    (b) $Pe$ dependence of $C_{\mathrm{B}}^{\mathrm{exp}}$, defined with respect to the Brownian stress.
    The solid and dashed lines represent $C_{\mathrm{B}}^{\mathrm{th}}$ obtained from the Fokker--Planck equation and $C_{\mathrm{B}}^{\mathrm{micro}}$ estimated from CNC material constants and microstructure, respectively.
    (c) Relationship between $C_{\mathrm{app}}/C_{\mathrm{B}}$ and the Brownian stress fraction.
    Experimental values were normalized by $\overline{C_{\mathrm{B}}^{\mathrm{exp}}}$, and the solid line represents $y=x$.}
    \label{fig:Capp_ratio}
\end{figure}

In the theoretical calculations in the previous section, the decrease in $C_{\rm app}$ in the high-$Pe$ regime occurred in the same $Pe$ range as the decrease in the Brownian stress contribution.
Here, this correspondence is examined using the experimental data.
Figure~\ref{fig:Capp_ratio}(a) shows $C_{\rm app}$ plotted against the Brownian stress contribution.

The experimental data obtained at different concentrations and $Pe$
numbers exhibited a roughly common trend when plotted against
$\tilde{\sigma}_{\rm B}/\sigma_{\rm Total}^{\rm exp}$.
This result is qualitatively consistent with the interpretation that the
concentration and $Pe$ dependences of $C_{\rm app}$ arise from changes in
the Brownian stress contribution to the total stress.
Although a systematic difference of approximately a factor of 1.9
remained between experiment and theory, the experimental data could be
approximately related to the theoretical prediction by a nearly common
rescaling factor over the concentrations and $Pe$ range examined here.
The approximate constancy of this factor suggests that the model captures
the principal concentration- and $Pe$-dependent trends associated with
stress partitioning, while a systematic difference remains in the
absolute scale.

As shown in Fig.~\ref{fig:Capp}(d), this difference is associated
mainly with the smaller values of
$\tilde{\sigma}_{\rm B}/\sigma_{\rm Total}^{\rm exp}$ relative to the
theoretical $\sigma_{\rm B}/\sigma_{\rm Total}$.
Possible sources of this difference include uncertainty in the absolute
value of the measured stress, the assumptions used to estimate
$\tilde{\sigma}_{\rm B}$, the choice of representative particle length,
and polydispersity-dependent differences in the optical and stress
weighting of the particle-length distribution.
Therefore, the factor of 1.9 should not be interpreted quantitatively as
a difference in the actual Brownian stress contribution.
Nevertheless, the approximately common concentration- and
$Pe$-dependent trend supports the interpretation that the decrease in
$C_{\rm app}$ is associated with a decrease in the Brownian stress
fraction of the total stress.

To further examine this correspondence, the stress--optic coefficient was evaluated using the Brownian stress, rather than the total stress.
For the experimental data, using the estimated Brownian stress $\tilde{\sigma}_{\rm B}$ introduced in the previous section, we defined
\begin{equation}
C_{\rm B}^{\rm exp}(Pe)=\frac{\Delta n \sin 2\chi}{2\tilde{\sigma}_{\rm B}} .
\end{equation}
Figure~\ref{fig:Capp_ratio}(b) shows the $Pe$ dependence of the obtained $C_{\rm B}^{\rm exp}$.
For comparison, the vertical-axis range was set to be the same as that for $C_{\rm app}$ in Fig.~\ref{fig:Capp}(a).
The theoretical value based on the Brownian stress $C_{\rm B}^{\rm th}$, as obtained from the Fokker--Planck equation, is also shown as a solid line.
The red dashed line represents $C_{\rm B}^{\rm micro}$, estimated from the CNC material constants and microstructure, as discussed below.

Whereas $C_{\rm app}$ varied substantially with concentration and $Pe$, the variation in $C_{\rm B}^{\rm exp}$ was relatively small.
The mean experimental value was $\overline{C_{\rm B}^{\rm exp}}=(4.25\pm0.69)\times10^{-5}~\mathrm{Pa^{-1}}$.
The theoretical value obtained from the Fokker--Planck equation was $C_{\rm B}^{\rm th}=2.24\times10^{-5}~\mathrm{Pa^{-1}}$, which is of the same order as the experimental value.
The difference corresponds to the systematic deviation between the experimental and theoretical values in Fig.~\ref{fig:Capp_ratio}(a), and could reflect uncertainty in the absolute value of $\tilde{\sigma}_{\rm B}$ estimated from the experimental data.
Nevertheless, under the stress-partitioning assumption adopted here, the concentration and $Pe$ dependences of $C_{\rm B}^{\rm exp}$ were much more strongly suppressed than those of $C_{\rm app}$.
This result is consistent with the SOR picture in which birefringence is more directly related to the Brownian stress than to the total stress \cite{Wagner1988,Bender1995,Bender1996}.

The relationship between $C_{\rm app}$ and $C_{\rm B}$ is now examined in terms of the Brownian stress contribution to the total stress.
From Eq.~\eqref{eq:CovsCapp}, we obtained the relationship $C_{\rm app}/C_{\rm B}=\sigma_{\rm B}/\sigma_{\rm Total}$.
Figure~\ref{fig:Capp_ratio}(c) shows $C_{\rm app}/C_{\rm B}$ plotted against the Brownian stress contribution.
For the experimental data, $\tilde{\sigma}_{\rm B}/\sigma_{\rm Total}^{\rm exp}$ was used as the horizontal-axis value, and $C_{\rm app}$ was normalized by the mean value $\overline{C_{\rm B}^{\rm exp}}$ obtained from Fig.~\ref{fig:Capp_ratio}(b).
The experimental data were distributed approximately along the line $y=x$.

This representation recasts the result in Fig.~\ref{fig:Capp_ratio}(b), where the variation of $C_{\rm B}^{\rm exp}$ with flow conditions was much smaller than that of $C_{\rm app}$, in terms of the relationship between the total-stress-based stress--optic coefficient and the Brownian stress contribution.
Thus, this scaling is consistent with Eq.~\eqref{eq:CovsCapp}, which describes the variation in $C_{\rm app}$ as arising from changes in the Brownian stress contribution to the total stress.
Based on this result, the deviation from the total-stress-based SOR in the high-$Pe$ regime does not necessarily indicate that the correspondence between birefringence and Brownian stress is lost.
Rather, it can be interpreted as an apparent change in $C_{\rm app}$ defined with respect to the total stress, caused by the increasing relative contributions of the solvent viscous stress and hydrodynamic stress compared with the Brownian stress.
However, $\tilde{\sigma}_{\rm B}/\sigma_{\rm Total}^{\rm exp}$ used as the horizontal-axis value is an estimated quantity obtained under the assumptions in Eq.~\eqref{eq:sigmaB_estimated}.
In particular, in the high-$Pe$ regime, stress partitioning is assumed based on the low-$Pe$ limit, and therefore its absolute value cannot be quantitatively interpreted as the true $\sigma_{\rm B}/\sigma_{\rm Total}$.
Thus, Fig.~\ref{fig:Capp_ratio}(c) should not be regarded as an independent quantitative validation of Eq.~\eqref{eq:CovsCapp}, but as an indication that the variation of $C_{\rm app}$ with the estimated Brownian stress contribution is consistent with the SOR interpretation based on the stress-component ratio.

The relationship between the total-stress-based SOR and the Brownian-stress-based relation can thus be understood through the Brownian stress contribution.
In the high-$Pe$ regime, $C_{\rm app}$ cannot be treated as a constant stress--optic coefficient.
However, its decrease does not necessarily indicate a complete breakdown of the stress--birefringence correspondence, but should be interpreted as an apparent response arising from the decrease in the Brownian stress contribution to the total stress.
Therefore, even when the total-stress-based SOR deviates from a simple proportional relationship, the stress--birefringence correspondence may be extended by considering the relative contributions of the individual stress components.

\subsection{Microscopic Estimate of $C_{\rm B}$}
We now discuss the magnitude of the obtained $C_{\rm B}$ from the viewpoint of the optical anisotropy and microstructure of the CNCs.
For this evaluation, the intrinsic birefringence $\Delta n_{0}=0.050$, determined from the comparison with the experiments in the previous section, was used to estimate the anisotropic polarizability of the anhydroglucose unit (AGU) constituting the CNCs.

Using the Lorentz--Lorenz equation, which relates anisotropic polarizability to refractive index \cite{Lorentz1881}, the intrinsic birefringence $\Delta n_0$ can be expressed in terms of the anisotropic polarizability per AGU, $\Delta\alpha_{\rm AGU}$ [m$^3$], as
\begin{equation}
\Delta n_0
=
\frac{2\pi}{9}
\frac{\rho_{\rm CNC} N_{\rm A}}{M_{\rm AGU}}
\frac{\left(\bar n^2+2\right)^2}{\bar n}
\Delta\alpha_{\rm AGU}.
\end{equation}
Here, $\rho_{\rm CNC}$ [kg/m$^3$] is the CNC density, $N_{\rm A}$ [mol$^{-1}$] is Avogadro's constant, $M_{\rm AGU}$ [kg/mol] is the molar mass of an AGU, and $\bar n$ is the average refractive index.
Using $\rho_{\rm CNC}=1,500~\mathrm{kg/m^3}$ (based on the manufacturer specification) and the value of $\bar n$ listed in Table~\ref{tab:Quantities}, the anisotropic polarizability per AGU was estimated to be $\Delta\alpha_{\rm AGU}=1.15\times10^{-30}~\mathrm{m^3}$.
To convert $\Delta\alpha_{\rm AGU}$ into the anisotropic polarizability per CNC rod, the number of AGUs contained in a single CNC rod was estimated.
Approximating a CNC rod as a cylinder, the number of AGUs per rod, $N_{\rm AGU}$, was estimated by dividing the rod volume by the volume occupied by one AGU in the crystal:
\[N_{\rm AGU}
=\frac{\pi d_n^2L_w/4}{A_{\rm AGU}l_{\rm AGU}}\approx3.4\times10^4.
\]
Here, $A_{\rm AGU}$ [m$^2$] is the cross-sectional area per AGU and $l_{\rm AGU}$ [m] is the length along the chain axis per AGU.
These values were taken as $0.3~\mathrm{nm^2}$ and $0.5~\mathrm{nm}$, respectively \cite{Hadden2014,Tashiro2019}.
Assuming that the anisotropic polarizability of the AGUs is additive, the anisotropic polarizability per CNC rod was evaluated as $\Delta\alpha_{\rm rod}=N_{\rm AGU}\Delta\alpha_{\rm AGU}$.
Under this assumption, $C_{\rm B}^{\rm micro}$ can be expressed as \cite{Doi1986}
\begin{equation}
C_{\rm B}^{\rm micro}
=
\frac{2\pi}{27k_{\rm b}T}
\frac{\left(\bar n ^2+2\right)^2}{\bar n}
\Delta\alpha_{\rm rod}.
\label{eq:C0}
\end{equation}
Substituting the material constants and rod dimensions obtained in this study yielded $C_{\rm B}^{\rm micro}=2.61\times10^{-5}~\mathrm{Pa^{-1}}$.

This microstructure-based estimate is close to $C_{\mathrm{B}}^{\mathrm{th}}$ obtained from the Fokker--Planck equation and is also of the same order as $\overline{C_{\mathrm{B}}^{\mathrm{exp}}}$ obtained from the stress measurements, as shown by the red dashed line in Fig.~\ref{fig:Capp_ratio}(b).
This consistency indicates that the experimentally obtained optical anisotropy of the CNCs can be connected, through their molecular structure and rod-like geometry, to the magnitude of the Brownian-stress-based stress--optic coefficient.
However, because the experimentally determined $\Delta n_{0}$ was used to evaluate $C_{\mathrm{B}}^{\mathrm{micro}}$, this comparison is not a fully independent validation of $C_{\mathrm{B}}$.
Rather, it demonstrates consistency among the optical response, microstructure, and stress response.

\section{Conclusion}
In this study, we investigated the relationship between flow-induced birefringence and shear stress in suspensions of rod-like cellulose nanocrystals using simultaneous rheo-optical measurements.
By combining the experiments with a Fokker--Planck-type description of orientational dynamics and a constitutive stress model, we examined how the apparent SOR based on the total stress is connected to the Brownian-stress-based optical response.

In the low-$Pe$ regime, birefringence and total stress exhibited an apparently linear SOR-type relation at each concentration.
However, the proportionality varied with concentration, indicating that the total-stress-based response reflects the relative contributions of different stress components rather than a single material relation.
When the optical response was instead referenced to an estimated Brownian-stress contribution, its dependence on concentration and flow conditions was substantially reduced.
This is consistent with birefringence being more directly associated with orientational Brownian stress than with the total stress.

At higher $Pe$, nonlinear flow-induced orientation and orientational saturation weakened the increase in birefringence.
The Fokker--Planck-based stress analysis showed that the relative Brownian contribution to the total stress decreases in this regime, while solvent viscous and hydrodynamic contributions remain significant.
Experimentally estimated Brownian-stress fractions showed the same overall decreasing trend.
Thus, the apparent deviation from the SOR at high $Pe$ does not necessarily imply a loss of the underlying orientation--stress correspondence, but can instead be interpreted as a consequence of changes in stress partitioning.

Overall, this study has provided a framework for interpreting the stress--optic response of rod-like suspensions in terms of stress partitioning.
By linking flow-induced orientational dynamics and optical anisotropy to macroscopic stress, this framework connects microscopic orientational behavior to the rheological response in anisotropic macromolecular and colloidal systems.

\section*{Nomenclature}
\begingroup
\setlength{\baselineskip}{10pt}
\subsection*{Roman symbols}
\begin{description}
\item[$C_{\rm app}$] Apparent stress--optic coefficient 
\hfill $\mathrm{Pa^{-1}}$
\item[$C_{\rm B}$] Brownian-stress-based stress--optic coefficient 
\hfill $\mathrm{Pa^{-1}}$
\item[$C_{\rm B}^{\rm exp}$] Experimentally estimated Brownian-stress-based stress--optic coefficient 
\hfill $\mathrm{Pa^{-1}}$
\item[$C_{\rm B}^{\rm th}$] Theoretical Brownian-stress-based stress--optic coefficient 
\hfill $\mathrm{Pa^{-1}}$
\item[$C_{\rm B}^{\rm micro}$] Stress--optic coefficient estimated from CNC optical anisotropy and microstructure
\hfill $\mathrm{Pa^{-1}}$
\item[$D_r^{0}$] Dilute-limit rotational diffusion coefficient 
\hfill $\mathrm{s^{-1}}$
\item[$D_r^{\rm eff}$] Effective rotational diffusion coefficient 
\hfill $\mathrm{s^{-1}}$
\item[$L_n$] Number-average particle length 
\hfill $\mathrm{m}$
\item[$L_w$] Weight-average particle length 
\hfill $\mathrm{m}$
\item[$Pe$] Effective Péclet number, $Pe=|\dot{\gamma}|/D_r^{\rm eff}$ 
\hfill -
\item[$Pe^0$] Bare Péclet number, $Pe^0=|\dot{\gamma}|/D_r^{0}$ 
\hfill -
\item[$S$] In-plane scalar order parameter 
\hfill -
\end{description}

\subsection*{Greek symbols}
\begin{description}
\item[$\beta$] Stretching exponent of the KWW function 
\hfill -
\item[$\Delta n$] Flow-induced birefringence 
\hfill -
\item[$\Delta n_0$] Intrinsic birefringence of CNC
\hfill -
\item[$\dot{\gamma}$] Shear rate 
\hfill $\mathrm{s^{-1}}$
\item[$\eta_{\rm s}$] Solvent viscosity 
\hfill $\mathrm{Pa}$ $\cdot$ ${\mathrm s}$
\item[$\nu_e$] Effective number density 
\hfill $\mathrm{m^{-3}}$
\item[$\psi$] CNC volume fraction 
\hfill -
\item[$\chi$] Orientation angle
\hfill $\mathrm{deg}$
\item[$\sigma_{\rm Total}$] Total shear stress 
\hfill $\mathrm{Pa}$
\item[$\sigma_{\rm S}$] Solvent viscous stress 
\hfill $\mathrm{Pa}$
\item[$\sigma_{\rm B}$] Brownian stress 
\hfill $\mathrm{Pa}$
\item[$\tilde{\sigma}_{\rm B}$] Estimated Brownian stress 
\hfill $\mathrm{Pa}$
\item[$\sigma_{\rm EV}$] Excluded-volume stress contribution 
\hfill $\mathrm{Pa}$
\item[$\sigma_{\rm H}$] Hydrodynamic stress contribution 
\hfill $\mathrm{Pa}$
\item[$\bar{\tau}$] Mean birefringence relaxation time 
\hfill $\mathrm{s}$
\end{description}

\section*{Acknowledgments}
The authors thank Assoc. Prof. Teshima of Kyushu University for valuable guidance on the use of atomic force microscopy.
This work was supported by JSPS KAKENHI Grant No. JP25KJ1211 and JST PRESTO Grant No. JPMJPR21O5.

\section*{Author Contributions}
{\bf William Kai Alexander Worby}: Conceptualization, Methodology, Investigation, Formal analysis, Funding acquisition, Writing – original draft.
{\bf Yuto Yokoyama}: Interpretation of results, Writing – review \& editing.
{\bf Misa Kawaguchi}: Interpretation of results, Writing – review \& editing.
{\bf Yoshiyuki Tagawa}: Supervision, Project administration, Funding acquisition, Writing – review \& editing.

All authors discussed the results and contributed to the final manuscript.

\section*{Data Availability}
Datasets generated during the current study are available from the corresponding author (YT) upon reasonable request.

\section*{Declaration of Competing Interest}
The authors declare that they have no competing financial interests or personal relationships that could have influenced the research presented in this manuscript.

\section*{Generative AI Statement}
Generative artificial intelligence tools (ChatGPT, OpenAI) were used during the preparation of this manuscript to enhance the clarity and quality of the language. 
All AI-assisted content was subsequently reviewed and revised by the authors, who assume full responsibility for the final published content.

\section{Appendix A: AFM Measurements}
\subsection{A1. Sample Preparation}
To evaluate the morphology of the CNCs, atomic force microscopy (AFM)
measurements were performed.
The samples were prepared by adsorbing CNCs onto
poly-L-lysine-treated mica substrates, following a procedure based on
the Langmuir--Blodgett method \cite{Oliveira2022}.

Freshly cleaved mica substrates (50-D-12, NanoAndMore) were treated
with an aqueous poly-L-lysine solution (0.1 w/v\%, Sigma-Aldrich).
After standing for 10 min, the substrates were rinsed with distilled
water and dried at 45~$^\circ$C.
The CNC stock suspension was diluted with ultrapure water to a
concentration of approximately $1.0\times10^{-3}~\mathrm{wt.\%}$.
The lower half of each mica substrate was then immersed in the dilute
CNC suspension for 10--20 s to allow the CNCs to adsorb onto the
surface.
After immersion, the substrates were rinsed five times with ultrapure
water and dried at room temperature for 24 h before AFM observation.

\subsection{A2. AFM Imaging Conditions and Image Analysis}
AFM measurements were performed using an aluminum-backside-coated
cantilever (200AC-NA, OPUS by MikroMasch; nominal spring constant:
$9~\mathrm{N/m}$; nominal tip radius: $7~\mathrm{nm}$).
The measurements were conducted in tapping mode.
AFM images were acquired over an area of
$5~\mu\mathrm{m}\times5~\mu\mathrm{m}$ with a resolution of
$512\times512$ pixels and a scan rate of $1.90$--$2.07~\mathrm{Hz}$.
Images were obtained at multiple positions on the mica substrate,
mainly around the region corresponding to the air--liquid interface
during immersion.
The observation positions were selected to minimize local bias in
particle density within this region.

During the measurements, the setpoint was set to $0.85$--$0.95$,
the free-oscillation amplitude (OscAmplitude) was set to
$1.27$--$1.36~\mathrm{V}$, and the P-gain and I-gain used for feedback
control were set to 200--202 and 99, respectively.

The obtained AFM images were analyzed using the open-source image
analysis software \href{https://gwyddion.net/}{Gwyddion}.
Previous studies have reported average CNC diameters of approximately
$3$--$15~\mathrm{nm}$
\cite{Beck-Candanedo2005,Jakubek2018,Lane2023,Tanaka2015}.
In the present measurements, the nominal tip radius was
$7~\mathrm{nm}$, which is comparable to the CNC diameter.
Therefore, the finite tip geometry may broaden the apparent lateral
width in the AFM images, leading to an overestimation of the actual
particle diameter.
Accordingly, the CNC cross section was assumed to be circular, and
the particle height measured by AFM was used as the particle diameter.

\section{Appendix B: Validation of the Rheo-Optical Measurement Conditions}

\subsection{B1. Rheometer Validation}

In this study, a concentric-cylinder measurement system was constructed
by combining a stress-controlled rheometer (MCR302, Anton Paar) with
a rotating inner cylinder fabricated using a 3D printer
(Form $3+$, FormLabs).
To evaluate the accuracy of the shear stress measurements, preliminary
experiments were performed using silicone oils with known viscosities.
Silicone oils with nominal kinematic viscosities of $10$, $100$,
$300$, and $500~\mathrm{cSt}$ (KF-96 series, Shin-Etsu Chemical)
were used.

In a concentric-cylinder rheometer, systematic errors in the measured
stress may arise from the machining accuracy of the cylinders,
imperfect concentricity, and small eccentricities.
Therefore, the theoretical shear stress was calculated from the
analytical solution for steady Taylor--Couette flow and compared with
the experimentally measured stress.

Let $r_i$ and $r_o$ denote the radii of the inner and outer cylinders,
respectively, and let $\omega$ denote the angular velocity of the
inner cylinder.
The azimuthal velocity distribution $v_{\theta}(r)$ for a Newtonian
fluid is given by
\begin{equation}
v_{\theta}(r)=\frac{r_i^2 \omega}{r_o^2-r_i^2}
\left(\frac{r_o^2}{r}-r\right).
\end{equation}
Using this velocity distribution, the theoretical shear stress at
the inner cylinder wall, $\sigma_{\mathrm{true}}$, was calculated as
\begin{equation}
\sigma_{\mathrm{true}}=-\eta r_i
\left.
\frac{\partial}{\partial r}
\left(
\frac{v_{\theta}}{r}
\right)
\right|_{r=r_i},
\end{equation}
where $\eta$ is the viscosity of the silicone oil.
The viscosity was determined from the nominal kinematic viscosity
and density of each silicone oil.

As shown in Fig.~\ref{fig:Error}(a), the experimentally measured
stress $\sigma_{\mathrm{exp}}$ and the theoretical stress
$\sigma_{\mathrm{true}}$ exhibited a linear relationship over the
entire measurement range.

\begin{figure*}[tb!]
\centering
\includegraphics[width=1\linewidth]{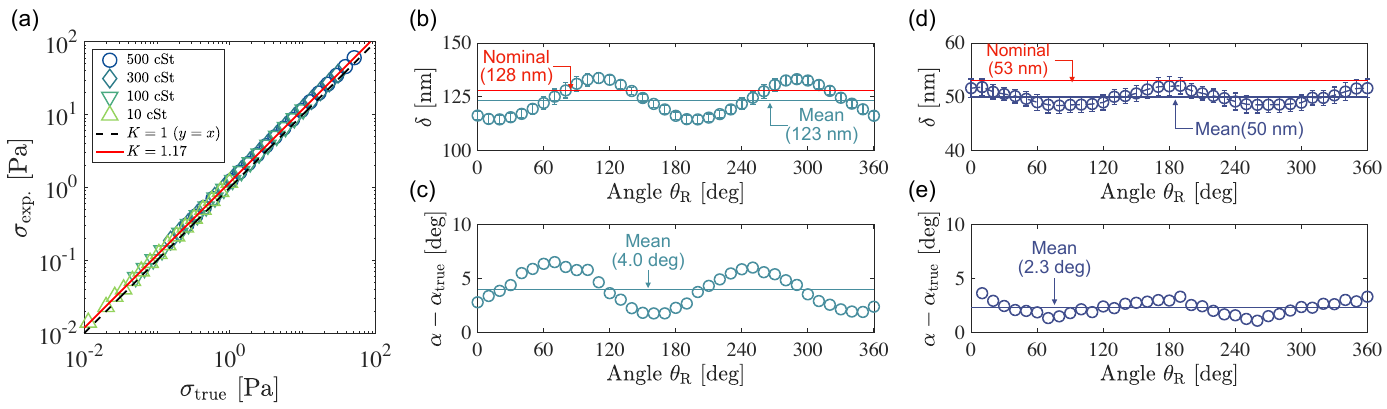}
\caption{\setstretch{0.8}
(a) Validation of the measurement accuracy of the concentric-cylinder
rheometer.
The theoretical shear stress $\sigma_{\mathrm{true}}$ and the
experimentally measured stress $\sigma_{\mathrm{exp}}$ exhibited a
linear relationship with a proportionality coefficient of $K=1.17$.
(b)--(e) Validation of the retardation and principal axis angle
measurements obtained using the polarization camera.
Panels (b) and (c) show the measured phase retardation $\delta$ and
the principal axis angle error $\alpha-\alpha_{\mathrm{true}}$,
respectively, as functions of the retarder rotation angle
$\theta_{\rm R}$ for a retarder with a nominal retardation of
$128~\mathrm{nm}$.
Panels (d) and (e) show the corresponding results for a retarder
with a nominal retardation of $53~\mathrm{nm}$.
In panels (b) and (d), the red lines indicate the nominal retardations,
and the solid lines indicate the mean measured values.
}
\label{fig:Error}
\end{figure*}

This relationship is expressed as
\begin{equation}
\sigma_{\mathrm{exp}}=K\sigma_{\mathrm{true}}.
\end{equation}
Linear fitting yielded a proportionality coefficient of $K=1.17$.
This result indicates that, under the present validation conditions,
the measured stress was approximately $17\%$ larger than the
theoretical value.
Although the absolute stress values obtained using this system may
therefore include a systematic uncertainty, the linear relationship
indicates that relative changes in stress can be evaluated with
reasonable reliability.

The stress values presented in the main text were not uniformly
corrected using this coefficient.
This is because $K$ was obtained from a preliminary scale check using
silicone oils and may not be directly applicable as a rigorous
calibration factor under all conditions of the suspension measurements.
Accordingly, the validation result was used as an indicator of the
systematic uncertainty in the absolute stress values associated with
the custom-made rheometer, rather than as a correction factor.
Although this scale difference should be considered in quantitative
comparisons, it does not substantially alter the interpretation of
the shear-rate (or $Pe$) and concentration dependences, which are the
main focus of this study.

\subsection{B2. Polarization Camera Validation}
To evaluate the accuracy of the retardation and principal axis angle
measurements obtained using the polarization camera
(CRYSTA PI-5WP, Photron), preliminary experiments were performed
using retarders with known retardations.
Each retarder was rotated by an angle $\theta_{\rm R}$, and the
retardation $\delta$ and principal axis angle $\alpha$ were measured
at each angle.
Here, $\theta_{\rm R}$ denotes the retarder rotation angle, and
$\alpha$ denotes the principal axis angle obtained from the
polarization camera.
For the analysis, a $300~\mathrm{pix}\times350~\mathrm{pix}$ region
at the center of the field of view was selected as the region of
interest (ROI).
The values of $\delta$ and $\alpha$ were evaluated as spatial
averages within the ROI.

Ideally, the retardation $\delta$ of a retarder should be independent
of the rotation angle $\theta_{\rm R}$.
Therefore, any angular dependence of the measured $\delta$ may
reflect errors originating from the optical measurement system.
Figures~\ref{fig:Error}(b) and (d) show the measured retardation as
a function of $\theta_{\rm R}$.
The red lines indicate the nominal retardation of each retarder,
whereas the solid lines indicate the mean measured values.
The error bars represent the standard deviation within the ROI.
Figures~\ref{fig:Error}(b) and (d) correspond to retarders with
nominal retardations of $128$ and $53~\mathrm{nm}$, respectively.

The measured retardation exhibited periodic variations with
$\theta_{\rm R}$.
For the retarder with a nominal retardation of $128~\mathrm{nm}$,
the measured values varied by approximately $\pm 8$--$10~\mathrm{nm}$
around the mean value of $123~\mathrm{nm}$, corresponding to a
relative peak variation of approximately $6$--$8\%$.
For the retarder with a nominal retardation of $53~\mathrm{nm}$,
the variation was approximately $\pm 2$--$3~\mathrm{nm}$ around
the mean value of $50~\mathrm{nm}$, corresponding to a relative
peak variation of approximately $4$--$6\%$.
Because these variations occurred even though the retardation should
ideally remain constant, they are considered to reflect angle-dependent
systematic errors associated with the polarization camera and the
optical system.

For the uncertainty propagation described below, a representative
standard uncertainty was required rather than the maximum deviation.
Therefore, these periodic variations were interpreted in terms of
a typical RMS-level uncertainty.
Based on the validation results, the representative relative
uncertainty in the retardation measurement was estimated as $5\%$.

Figures~\ref{fig:Error}(c) and (e) show the difference between the
measured principal axis angle $\alpha$ and the expected angle
$\alpha_{\mathrm{true}}$ determined from the retarder setting angle.
The difference $\alpha-\alpha_{\mathrm{true}}$ exhibited periodic
variations with $\theta_{\rm R}$, indicating angle-dependent
systematic deviations associated with the polarization camera and
optical system.
The representative magnitude of these variations was estimated as
approximately $3^\circ$, corresponding to the median value obtained
from the two retarders.

It should be noted that the variations in $\delta$ and $\alpha$
obtained in this validation do not directly establish the absolute
accuracy of the retardation after the unwrapping procedure.
Rather, these results were used as indicators of the systematic
uncertainties associated with the polarization camera and optical
system within the non-wrapped measurement range.

\subsection{B3. Region-of-Interest Dependence}
To evaluate the sensitivity of the measured birefringence and
orientation angle to the selection of the ROI, its radial width was
systematically varied while keeping its center fixed at the midpoint
of the Taylor--Couette gap.
The ROI widths examined were 1, 3, 5, 11, 21, and 33 pixels.
The 33-pixel ROI corresponds approximately to the full radial width
of the flow gap, as illustrated in Fig.~\ref{fig:ROI}(a).

Figures~\ref{fig:ROI}(b) and (c) show the shear-rate dependence of
the spatially averaged birefringence $\langle\Delta n\rangle$ and
orientation angle $\langle\chi\rangle$, respectively, for the
different ROI widths.
The mean values of $\langle\Delta n\rangle$ were nearly independent
of the ROI width over the investigated shear-rate range.
Similarly, $\langle\chi\rangle$ exhibited only limited dependence
on the ROI width over most of the measurement range.
These results indicate that the characteristic shear-rate dependence
of the rheo-optical quantities is not strongly affected by the
specific ROI width used in the main analysis.

The error bars in Figs.~\ref{fig:ROI}(b) and (c) represent the
standard deviation of the measured values within each ROI.
They therefore characterize the spatial variation within the
observation region rather than the uncertainty in the spatially
averaged value.
The spatial variation tended to increase with increasing ROI width,
particularly for the 33-pixel ROI covering almost the entire flow gap.
This increase may arise, at least in part, from the inclusion of a
wider radial region and slight variations in the local gap position
and width associated with imperfect concentricity of the custom-made
Taylor--Couette geometry.

Nevertheless, the mean values and their overall shear-rate
dependences remained largely unchanged with ROI width.
Accordingly, a 3-pixel-wide ROI centered at the midpoint of the gap
was used in the main analysis to obtain representative rheo-optical
quantities while minimizing the influence of spatial variations
across the gap.

\begin{figure*}[tb!]
\centering
\includegraphics[width=1.0\linewidth]{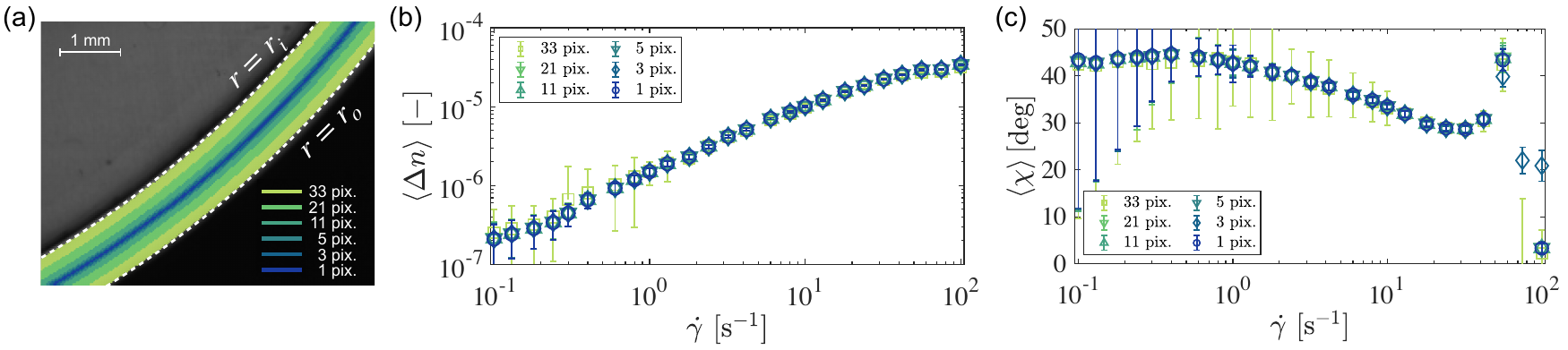}
\caption{\setstretch{0.8}
Effect of the ROI width on the measured rheo-optical quantities.
(a) Definition of the ROIs used for the sensitivity analysis.
All ROIs were centered at the midpoint of the Taylor--Couette gap,
and their radial widths were varied from 1 to 33 pixels.
(b) Spatially averaged birefringence $\langle\Delta n\rangle$ and
(c) orientation angle $\langle\chi\rangle$ as functions of shear
rate for the different ROI widths.
Symbols indicate the spatial mean within each ROI, and error bars
represent the standard deviation within the corresponding ROI.
}
\label{fig:ROI}
\end{figure*}

\subsection{B4. Phase Wrapping and Unwrapping Procedure}

Here, the phase wrapping and unwrapping procedures used in this study
are described.
Because optical retardation is periodic, when the actual retardation
exceeds the measurable range, the measured value is folded back to
a smaller value representing the same or an equivalent polarization
state \cite{Yokoyama2023,Muto2025c}.
This phenomenon is referred to as phase wrapping and must be
corrected to evaluate the absolute retardation quantitatively.

In the polarization camera used in this study (CRYSTA PI-5WP,
Photron), the retardation is output in the range
$0\leq\delta_{\mathrm{meas}}\leq\lambda/2$.
Therefore, when the actual retardation exceeds $\lambda/2$, the
measured value is folded back into this range.
Phase wrapping is also accompanied by an interchange of the slow
and fast axes, requiring correction of the measured principal axis
angle.
Accordingly, the unwrapping procedure was applied to both the
retardation $\delta$ and the principal axis angle $\alpha$.

The wrapping index $m$ was determined by assuming that the retardation
varies continuously with time or shear rate.
Here, $m$ is an integer representing the folding interval for each
$\lambda/2$.
Let $\delta_{\mathrm{meas}}$ and $\delta_{\mathrm{corr}}$ denote the
retardation before and after unwrapping, respectively.
The corrected retardation was calculated as
\begin{equation}
\delta_{\mathrm{corr}}
=
\begin{cases}
\dfrac{m}{2}\lambda+\delta_{\mathrm{meas}},
& m \ \mathrm{is~0~or\ even},\\
\dfrac{m+1}{2}\lambda-\delta_{\mathrm{meas}},
& m \ \mathrm{is\ odd}.
\end{cases}
\end{equation}
Here, $\lambda$ is the wavelength of the light used for measurement.
In the present experiments, $\lambda=540~\mathrm{nm}$.
This expression accounts for the reversal of the apparent direction
of change in the measured retardation each time wrapping occurs.

When $m$ is odd, a correction of $90^\circ$ was applied to the
measured principal axis angle $\alpha_{\mathrm{meas}}$.
The corrected principal axis angle $\alpha_{\mathrm{corr}}$ was
calculated as
\begin{equation}
\alpha_{\mathrm{corr}}
=
\begin{cases}
\alpha_{\mathrm{meas}},
& m \ \mathrm{is~0~or\ even},\\
\alpha_{\mathrm{meas}}-90^\circ,
& m \ \mathrm{is\ odd}.
\end{cases}
\end{equation}

The wrapping index $m$ was determined from the continuity of the
measured retardation between adjacent measurement points.
When $\delta_{\mathrm{meas}}$ exhibited a discontinuous change
with time or shear rate, the change was attributed to phase wrapping,
and $m$ was updated accordingly.

Figure~\ref{fig:Unwrap}(a) shows the time evolution of the
retardation before and after flow cessation.
Because the uncorrected retardation is folded into the range below
$\lambda/2$, it appears to increase immediately after flow cessation.
In contrast, the corrected retardation exhibits a continuous,
monotonic decay after cessation.
As shown in Fig.~\ref{fig:Unwrap}(b), the uncorrected and corrected
values coincide in the low-shear-rate regime under steady shear.
At higher shear rates, where the actual retardation exceeds
$\lambda/2$, the uncorrected value apparently decreases owing to
wrapping, whereas the corrected value exhibits a continuous
increasing trend.

It should be noted that the unwrapping procedure assumes continuity
of the retardation.
Therefore, additional uncertainty may arise in the corrected
retardation and principal axis angle near the wrapping points.
Although the corrected values were used in the subsequent analysis,
their absolute magnitudes should be regarded as including
uncertainties associated with the polarization camera, optical
system, and unwrapping procedure.

\begin{figure}[tb!]
\centering
\includegraphics[width=1\linewidth]{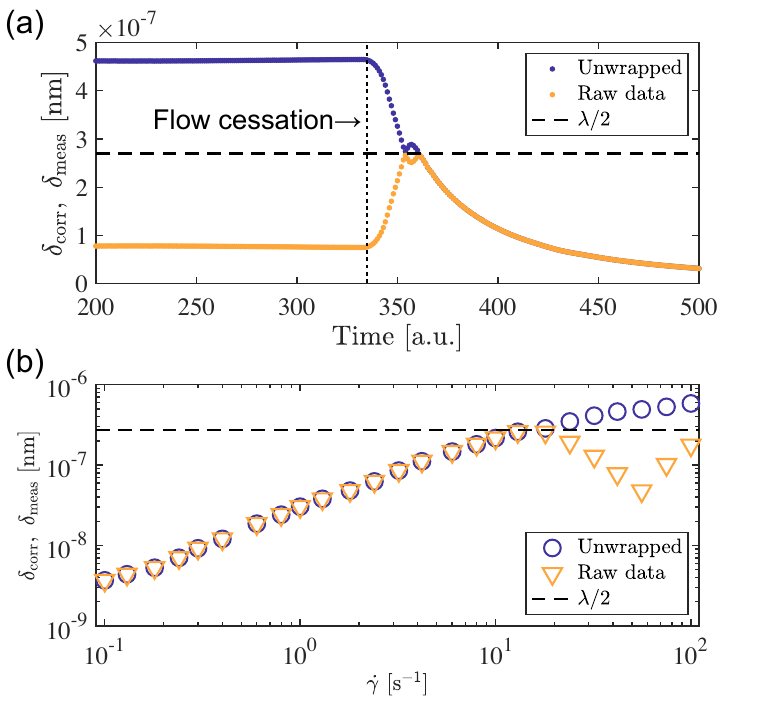}
\caption{\setstretch{0.8}
Phase unwrapping procedure for the measured retardation.
(a) Time evolution of the retardation before and after unwrapping
during flow cessation.
The vertical dashed line indicates the flow cessation time.
(b) Shear-rate dependence of the retardation before and after
unwrapping.
Purple circles and orange inverted triangles denote the corrected
and raw measured retardations, respectively.
The black dashed line represents $\lambda/2$, the upper measurable
limit of the polarization camera.
}
\label{fig:Unwrap}
\end{figure}

\subsection{B5. Uncertainty Estimation for the Stress-Optic Coefficient}

The stress--optic coefficient $C$ evaluated in this study includes
uncertainties associated with the measurements of birefringence,
principal axis angle, and shear stress.
These contributions were estimated using first-order error
propagation.

The stress--optic coefficient is defined as
$C=\Delta n\sin(2\chi)/(2\sigma)$.
Here, $\chi$ denotes the orientation angle characterizing the
optically measured anisotropy.
Therefore, the uncertainty in the principal axis angle was treated
as the uncertainty in $\chi$.
The standard uncertainties in $\Delta n$, $\chi$, and $\sigma$ are
denoted by $u_{\Delta n}$, $u_{\chi}$, and $u_{\sigma}$,
respectively, with $u_{\chi}$ expressed in radians.

The uncertainty in $C$ was estimated using the first-order law of
propagation of uncertainty described in the Guide to the Expression
of Uncertainty in Measurement (GUM)
\cite{Taylor1994,JCGMGUM}.
Assuming that the input quantities are uncorrelated, the combined
standard uncertainty is given by
\[
u_C^2
=
\left(
\frac{\partial C}{\partial\Delta n}
\right)^2u_{\Delta n}^2
+
\left(
\frac{\partial C}{\partial\chi}
\right)^2u_{\chi}^2
+
\left(
\frac{\partial C}{\partial\sigma}
\right)^2u_{\sigma}^2.
\]
The corresponding sensitivity coefficients are
\[
\begin{aligned}
\frac{\partial C}{\partial\Delta n}
&=
\frac{\sin(2\chi)}{2\sigma}
=
\frac{C}{\Delta n},
\\
\frac{\partial C}{\partial\chi}
&=
\frac{\Delta n\cos(2\chi)}{\sigma}
=
2C\cot(2\chi),
\\
\frac{\partial C}{\partial\sigma}
&=
-\frac{\Delta n\sin(2\chi)}{2\sigma^2}
=
-\frac{C}{\sigma}.
\end{aligned}
\]
Thus,
\begin{equation}
\left(\frac{u_C}{C}\right)^2
=
\left(\frac{u_{\Delta n}}{\Delta n}\right)^2
+
\left\{2\cot(2\chi)u_{\chi}\right\}^2
+
\left(\frac{u_{\sigma}}{\sigma}\right)^2.
\label{eq:Error}
\end{equation}

Strictly, however, the uncertainties in $\Delta n$ and $\chi$
may include common angle-dependent systematic errors because both
quantities are extracted from the same polarization image.
Since their covariance was not explicitly evaluated, the
uncorrelated form of the propagation law was used as an approximate
estimate of the representative uncertainty in $C$.

The birefringence was obtained from the measured retardation
$\delta$ and optical path length $H$ as $\Delta n=\delta/H$.
In this study, the uncertainty in $H$ was assumed to be sufficiently
small.
Accordingly, the relative uncertainty in the retardation measurement
associated with the polarization camera and optical system was adopted
as
\[
\frac{u_{\Delta n}}{\Delta n}
\simeq
\frac{u_{\delta}}{\delta}
\simeq0.05.
\]
From the validation experiments using retarders, the uncertainty in
the principal axis angle was estimated as
$u_{\chi}\simeq3^\circ=5.2\times10^{-2}~\mathrm{rad}$.
The representative relative uncertainty in the shear stress was
estimated as $u_{\sigma}/\sigma\simeq0.17$ from the preliminary
experiments using silicone oils.

The principal axis angle $\chi$ varied approximately from
$20^\circ$ to $45^\circ$ over the measurement range.
As shown in Fig.~\ref{fig:Uncertaintly}, the relative uncertainty
in $C$ estimated using Eq.~\eqref{eq:Error} was approximately
$17.7\%$ near $\chi=45^\circ$, $18.7\%$ at $\chi=30^\circ$,
and $21.5\%$ at $\chi=20^\circ$.
The representative uncertainty over the experimentally observed
range of $\chi$ was therefore estimated as
$\overline{u_C/C}\approx19\%$.

The dominant contribution to this estimate was the approximately
$17\%$ scale uncertainty in the stress measurement.
In contrast, the contribution from the principal axis angle
uncertainty was small when $\chi$ was close to $45^\circ$ but
increased as $\chi$ decreased.
Around $\chi\simeq10^\circ$, this contribution increased the
relative uncertainty in $C$ by several percent.

In addition, under conditions where phase wrapping occurred,
the corrected optical quantities may include uncertainties that
are not represented by Eq.~\eqref{eq:Error}.
Therefore, the above evaluation should be regarded as a local
estimate of the representative uncertainty for data points for
which the branch selection in the unwrapping procedure was judged
to be appropriate.

\begin{figure}[tb!]
\centering
\includegraphics[width=1\linewidth]{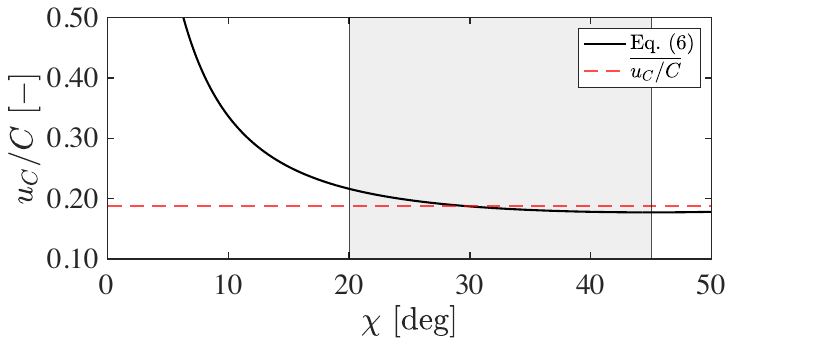}
\caption{\setstretch{0.8}
Relative uncertainty in the stress--optic coefficient $C$ as a
function of $\chi$.
The vertical solid lines indicate the experimentally observed
range of $\chi$, approximately
$20^\circ\leq\chi\leq45^\circ$.
The red dashed line represents the average relative uncertainty
over this range, $\overline{u_C/C}\simeq0.19$.
}
\label{fig:Uncertaintly}
\end{figure}

\section{Appendix C: Assessment of Flow Conditions}

\subsection{C1. Flow Stability in the Taylor--Couette Cell}

To assess the possibility of centrifugal instability in the present
Taylor--Couette measurements, the Taylor number was evaluated
following Khirennas et al. \cite{Khirennas2021}.
They defined the Taylor number $Ta$ for a system with a rotating
inner cylinder and a stationary outer cylinder as
\[
Ta=
\frac{\omega r_i(r_o-r_i)}
{\mu\sqrt{(r_o-r_i)/r_i}},
\]
where $\mu~[\mathrm{m^2/s}]$ is the kinematic viscosity of the fluid.

In the present study, the solvent viscosity and density were used
as representative material properties of the CNC suspension, with
$\eta_s=0.240~\mathrm{Pa}\cdot\mathrm{s}$ and
$\rho_s=1.23\times10^3~\mathrm{kg/m^3}$, respectively.
The kinematic viscosity was therefore estimated as
$\mu=\eta_s/\rho_s=1.95\times10^{-4}~\mathrm{m^2/s}$.

The maximum angular velocity employed in the measurements was
$\omega_{\max}=5.45~\mathrm{rad/s}$, yielding a maximum Taylor
number of $Ta_{\max}=2.13$.

Khirennas et al. \cite{Khirennas2021} reported the onset of Taylor
vortex flow at a critical Taylor number of $Ta_{c1}=43.8$ for a
Taylor--Couette system with a radius ratio
$r_i/r_o\simeq0.90$ and an aspect ratio
$H/(r_o-r_i)=15$, where $H$ is the length of the rotating
inner cylinder.
For comparison, the corresponding geometrical parameters of the
present setup were $r_i/r_o=0.95$ and $H/(r_o-r_i)=10$.
Although the geometries are not identical, the maximum Taylor
number in the present measurements was more than an order of
magnitude smaller than the reported critical value.
This comparison indicates that the present measurements were
performed well below the expected onset of Taylor vortex flow.

Khirennas et al. also showed that weak endwall-induced Ekman
structures can appear at Taylor numbers substantially below the
onset of Taylor vortex flow.
Therefore, the above comparison is used specifically to assess the
possibility of centrifugal instability leading to Taylor vortex flow,
rather than to assume ideal circular Couette flow throughout the
finite-length measurement cell.

\subsection{C2. Flow-Decay Timescale after Cessation}

The characteristic decay time of the flow after cessation in plane
Couette flow with a gap width $h$ is given by
\cite{batchelor1967}
\begin{equation}
\tau_f=\frac{h^2}{\pi^2\mu},
\end{equation}
where $\mu$ is the kinematic viscosity.

In the present system,
$\mu=1.95\times10^{-4}~\mathrm{m^2/s}$.
Using the gap width
$h\equiv r_o-r_i=1.0\times10^{-3}~\mathrm{m}$,
the characteristic flow-decay time was estimated as
$\tau_f=5.2\times10^{-4}~\mathrm{s}$.
Thus, the imposed shear flow decays on a sub-millisecond timescale
after cessation.

Figure~\ref{fig:Cessation} shows the ratio of the mean
birefringence relaxation time $\bar{\tau}$ obtained from KWW
fitting to the flow-decay time $\tau_f$.
The values of $\bar{\tau}/\tau_f$ obtained in this study were
approximately in the range
$1.0\times10^2$--$2.5\times10^2$.
Therefore, the characteristic decay time of the flow field was
at least two orders of magnitude shorter than the observed
birefringence relaxation time.

Following the timescale argument used by Yokoyama et al.
\cite{Yokoyama2026}, cessation of the imposed flow can therefore
be regarded as quasi-instantaneous on the timescale of orientational
relaxation.
Accordingly, the subsequent birefringence decay is interpreted
predominantly as the relaxation of CNC orientation under effectively
stopped-flow conditions.

In general, if the orientational relaxation timescale becomes
comparable to $\tau_f$, the measured birefringence decay may
contain contributions from both orientational relaxation and the
finite hydrodynamic decay of the shear flow.
In the present measurements, however, the large separation between
$\bar{\tau}$ and $\tau_f$ suggests that such coupling is negligible.
The analysis of the flow-cessation experiments therefore assumes
effectively instantaneous cessation of the shear flow relative to
the rotational relaxation of the CNCs.

\begin{figure}[tb!]
\centering
\includegraphics[width=1\linewidth]{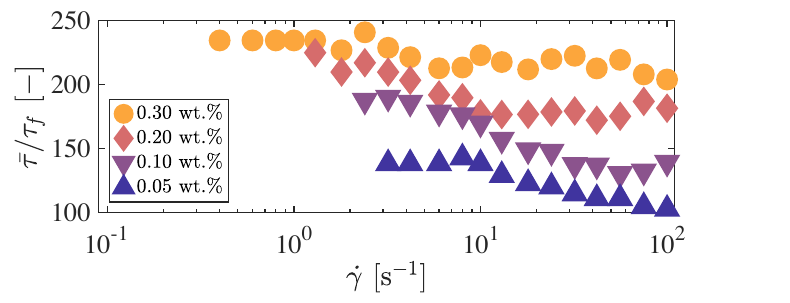}
\caption{\setstretch{0.8}
Ratio of the mean birefringence relaxation time $\bar{\tau}$
obtained from KWW fitting to the characteristic flow-decay time
$\tau_f$ after cessation.
The horizontal axis represents the steady shear rate applied
before flow cessation.
For all conditions under which reliable KWW fitting was possible,
$\bar{\tau}/\tau_f$ was of $O(10^2)$ or larger, indicating that
the flow field decays substantially faster than the orientational
relaxation of the CNCs.
}
\label{fig:Cessation}
\end{figure}

\bibliography{Ref}
\end{document}